\documentclass{article}
\usepackage{slashed,verbatim,subfigure}
\usepackage[numbers,sort&compress]{natbib}
\usepackage{amsmath}
\usepackage{amssymb}
\usepackage{authblk}
\usepackage{appendix}
\usepackage{graphicx}
\usepackage{dcolumn}
\usepackage{bm}
\usepackage[colorlinks=true,linktocpage=true,
linkcolor=blue,citecolor=blue]{hyperref}
\usepackage[a4paper]{geometry}
\newcommand{\be}{\begin{eqnarray}}
\newcommand{\ee}{\end{eqnarray}}
\newcommand{\ec}{\sigma_{\rm el}}

\begin{document}
\large
\title{\bf{Anisotropic modifications to the transport phenomena and observables in a hot QCD medium at finite baryon asymmetry}}
\author{Shubhalaxmi Rath\thanks{shubhalaxmirath@gmail.com}}
\affil{Centro Multidisciplinario de F\'isica, Vicerrector\'ia de Investigaci\'on, Universidad Mayor, 8580745 Santiago, Chile}
\date{}
\maketitle

\begin{abstract}
We have studied how the transport of charge and heat as well as associated observables become influenced by a weak-momentum anisotropy arising due to the asymptotic expansion of baryon asymmetric matter in the 
initial stages of ultrarelativistic heavy ion collisions. This study facilitates the understanding of the local equilibrium property of the medium through the Knudsen number, and explores the correlation between the heat flow and the charge flow through the Lorenz number in the Wiedemann-Franz law for an anisotropic hot QCD medium at finite baryon asymmetry. We have determined the electrical and the thermal conductivities by solving the relativistic Boltzmann transport equation in the relaxation time approximation within the kinetic theory approach. The interactions among partons are appended through their distribution functions within the quasiparticle model of the hot QCD medium at finite temperature, anisotropy and baryon asymmetry. We have observed a decrease in both electrical conductivity and thermal conductivity in the presence of expansion-induced anisotropy for baryonless scenario as well as for baryon asymmetric scenario. Conversely, these conductivities are found to be larger in the baryon asymmetric matter as compared to their counterparts in the baryonless matter. The impact of anisotropy on the baryon asymmetric matter is observed to be as conspicuous as on the baryonless matter. The above results are broadly attributed to two factors: the squeezing of the distribution functions due to the momentum anisotropy generated by the asymptotic expansion of baryon asymmetric matter and the dispersion relations of partons in the presence of anisotropy. Additionally, the aforesaid observables are also significantly modulated by the expansion-induced anisotropy in the baryon asymmetric medium, indicating new predictions for the equilibrium characteristic and the relative behavior between the heat and charge flow for the said medium. 

\end{abstract}

\newpage

\section{Introduction}
Ultrarelativistic heavy ion collisions at Relativistic Heavy Ion Collider (RHIC) and Large Hadron Collider (LHC) create a new state of the strongly interacting matter, known as quark-gluon plasma (QGP) and 
have been successfully collecting the evidences in the form of dilepton and photon spectra \cite{Feinberg:NCA34'1976,Shuryak:PLB78'1978,Kapusta:PRD44'1991}, anomalous quarkonium suppression \cite{Blaizot:PRL77'1996,Satz:NPA783'2007,Rapp:PPNP65'2010}, elliptic flow \cite{Bhalerao:PLB641'2006,Voloshin:PLB659'2008}, jet quenching \cite{Wang:PRL68'1992,Adcox:PRL88'2002,Chatrchyan:PRC84'2011} etc. for the production of such a matter. The initial stages of ultrarelativistic heavy ion collisions may exhibit a small but finite baryon asymmetry. According to some studies, at temperature around 160 MeV, the baryon chemical potential is approximately 300 MeV \cite{P:JPG28'2002,Cleymans:JPG35'2008,Andronic:NPA837'2010}. Additionally, in the strong magnetic field regime, the baryon chemical potential was observed to go up from 0.1 GeV to 0.6 GeV \cite{Fukushima:PRL117'2016}, implying an increase in the quark chemical potential. Various transport coefficients of the partonic medium were found to be noticeably modulated by the cumulative effects of magnetic field and baryon asymmetry \cite{Rath:EPJC80'2020,Rath:EPJC81'2021,Rath:EPJC82'2022,Rath:EPJA59'2023}. The presence of baryon asymmetry in chiral systems at finite temperature may induce axial currents \cite{Pu:PRD91'2015,Gorbar:PRD93'2016}. In the early stages of ultrarelativistic heavy ion collisions, a momentum anisotropy can also arise in the local rest frame of the fireball due to the asymptotically free expansion of fireball along the beam direction as compared to its transverse direction \cite{Dumitru:PLB662'2008,Dumitru:PRD79'2009}. This motivates a detailed exploration of the transport coefficients in a more realistic medium that incorporates both momentum-space anisotropy and finite baryon asymmetry. The anisotropy can be quantified by an anisotropic parameter ($\xi$) which is associated with the transverse ($p_T$) and longitudinal ($p_L$) components of momentum. In this type of anisotropy, the transverse component of momentum exceeds the longitudinal component of momentum, resulting in a consistently positive anisotropic parameter ($\xi=\langle {\bf p}_{T}^{2}\rangle/(2\langle p_{L}^{2}\rangle)-1$). If the anisotropy is weak ($\xi<1$), the parton distribution can be approximated by compressing the isotropic distribution along a specific direction, and the influence of such anisotropic distributions has been extensively investigated recently through various phenomenological and theoretical observations. For instance, the leading-order dilepton and photon yields were found to increase as a result of anisotropy \cite{Martinez:PRC78'2008,Ryblewski:PRD92'2015,Mukherjee:EPJA53'2017,Bhattacharya:PRD93'2016}. The emergence of anisotropy has been found to enhance the binding energies of heavy quark bound states \cite{Dumitru:PRD79'2009}. Additionally, the heavy quarkonium was observed to dissociate earlier in an anisotropic medium than its counterpart in an isotropic medium \cite{Thakur:PRD88'2013}. Further, the electrical conductivity was found to decrease with the increase of weak-momentum anisotropy \cite{Srivastava:PRC91'2015}. Conversely, the electrical and the thermal conductivities were observed to increase due to the presence of a strong magnetic field-induced anisotropy \cite{Rath:PRD100'2019}. In earlier works on the transport coefficients, the thermal masses of partons were assumed to be unaffected by the expansion-induced anisotropy. In this work, we consider the influence of anisotropy on the parton masses, alongside the effects of temperature and baryon asymmetry. Accordingly, we calculate the quasiparticle masses of partons as functions of the temperature, chemical potential and anisotropic parameter within the quasiparticle model, and utilize them to study different transport coefficients and observables in an expansion-induced anisotropic hot QCD medium with finite baryon asymmetry. 

Charge and heat transport processes in a medium can be well understood by studying the corresponding transport coefficients for that medium, such as the electrical conductivity ($\sigma_{\rm{el}}$) and the thermal conductivity ($\kappa$). Electrical conductivity characterizes the linear response of the medium to an external electric field and plays a significant role in modulating various processes in the early universe as well as in ultrarelativistic heavy ion collisions. Intense electric fields ($eE\simeq m_\pi^2$, with $m_\pi$ being the pion mass) can be generated during the early stages of ultrarelativistic heavy ion collisions \cite{Tuchin:AHEP2013'2013,Hirono:PRC90'2014}. In mass asymmetric collisions, the electric field tends to have an overall preferred direction, leading to a charge asymmetric flow and the strength of the flow is associated with the electrical conductivity of the medium \cite{Hirono:PRC90'2014}. The current induced in such collisions results in a dipole deformation of the charge distribution within the medium. The production rate 
of the thermal dileptons can be expressed in terms of the electric current-current correlation function \cite{McLerran:PRD31'1985,Gale:PRC35'1987,Ding:PRD83'2011}. Since its small frequency region is governed by the transport peak \cite{Forster:BOOK'1990}, the electrical conductivity can be estimated through the comparison 
of the theoretical results with the dilepton invariant mass spectra \cite{Akamatsu:PRC85'2012}. The electrical conductivity of the QGP under various conditions is essential in observing the lifetimes of the strong magnetic fields produced in the noncentral events of ultrarelativistic heavy ion collisions \cite{Tuchin:AHEP2013'2013,McLerran:NPA929'2014,Huang:RPP79'2016,Rath:PRD100'2019,Rath:EPJA59'2023,
Rath:EPJC83'2023}. Furthermore, electrical conductivity is used as an essential input for many phenomenological applications in RHIC and LHC, for example, the emission rate of soft photons \cite{Kapusta:BOOK'2006}. Additionally, the study of the thermal conductivity is essential to understand the efficiency of heat flow or the energy dissipation in a thermal medium. In the nonrelativistic case, the heat equation is obtained by the validity of the first and the second laws of thermodynamics, where the flow of heat is proportional to the temperature gradient with the proportionality factor being the thermal conductivity. The heat does not flow directly, instead, it diffuses depending on the internal structure of the medium it travels through. Similarly, in a relativistic QCD system, the heat flow depends on the intrinsic properties of the medium. It can also be affected by different external conditions as well as by anisotropy. Thermal conductivity may leave significant imprints on the hydrodynamic evolution of the medium with finite baryon asymmetry. In this work, we intend to explore the impact of anisotropy on the electrical and the thermal conductivities by calculating them in the presence of weak-momentum anisotropy caused by the asymptotic expansion of matter. We use the relativistic Boltzmann transport equation with the relaxation time approximation to determine the aforesaid conductivities for the expansion-induced anisotropic hot QCD medium at finite baryon asymmetry. We also aim to compare the effect of anisotropy on baryon asymmetric matter with that on baryonless matter, which would enable to perceive the role of baryon asymmetry in an anisotropic medium. The interactions among partons are 
incorporated through their distribution functions within the quasiparticle model. 

We further intend to study the observables associated with the charge and heat transport properties, such 
as the Knudsen number and the Lorenz number for a baryon asymmetric matter in the presence of 
expansion-induced anisotropy. The Knudsen number ($\Omega$) is essential to understand the equilibrium property of the medium as it is the ratio of the mean free path ($\lambda$) to the characteristic length of the system, where $\lambda$ is related to the thermal conductivity ($\lambda=3\kappa/(vC_V)$, where $v$ and $C_V$ are the relative speed and the specific heat at constant volume, respectively). The validity of equilibrium is satisfied, if the mean free path is much smaller than the characteristic length of the system, {\em i.e.} the medium can approach towards the equilibrium state, if there is sufficient separation between the microscopic and macroscopic length scales of the medium. In this work, our intention is to understand how the expansion-induced anisotropy affects the Knudsen number of a baryon asymmetric hot QCD medium. The electronic contributions of the thermal and electrical conductivities are related to each other as their ratio is 
equal to the product of the Lorenz number ($L$) and the temperature, widely known as the Wiedemann-Franz law. In fact, the ratio, $\kappa/\sigma_{\rm el}$ has approximately the same value for different metals at the 
same temperature. In the metallic phase, the electronic contribution to the thermal conductivity is much smaller than what would be expected from the Wiedemann-Franz law, which can be attributed to the independent propagation of charges and heat in a strongly correlated system. In this work, we intend to observe 
how the Lorenz number of a baryon asymmetric hot QCD medium gets affected by the anisotropy generated 
due to the asymptotic expansion of matter. 

The transport properties of the hot QCD medium and associated observables have been extensively studied 
within the relaxation time approximation and quasiparticle approaches under various physical 
scenarios. Earlier formulations of the framework used in this work were developed in references \cite{Rath:PRD100'2019,Rath:EPJC80'2020}. Subsequent works, such as references \cite{Khan:PRD104'2021,Gowthama:PRD106'2022,Rath:EPJA59'2023,Rath:EPJC83'2023,Shaikh:PRD108'2023,
Singh:PRD109'2024}, extended these ideas to investigate the charge and the heat transport properties in the presence of the different strengths of magnetic fields and/or the time-dependent magnetic fields. While these works provide important insights into magnetic field-driven transport phenomena, the present study addresses a fundamentally different physical setting, where we focus on the collective effects of expansion-induced momentum anisotropy and baryon asymmetry on the transport coefficients and observables of the hot QCD medium 
in the absence of magnetic field. This distinguishes our objectives and physical assumptions from the abovementioned magnetized-medium analyses and highlights the complementary nature of the present work. 

In brief, we have observed that the electrical and the thermal conductivities of the baryon asymmetric hot QCD medium get reduced in the presence of expansion-induced anisotropy as compared to their counterparts 
in the isotropic medium. In an anisotropic environment, these conductivities consist of an isotropic part and an additional anisotropic part. Anisotropic parts of the abovementioned conductivities contribute in reducing the conduction of charge and heat in the anisotropic medium. However, the baryon asymmetric medium estimates larger values of these conductivities as compared to the baryonless medium. This increase in $\ec$ and $\kappa$ is chiefly due to the enhanced parton number densities in the baryon asymmetric medium. Further, we have observed that the emergence of expansion-induced anisotropy makes the Knudsen number smaller than its value in the isotropic medium, thus pushing the medium closer to the local equilibrium state, whereas a slight increase of the Knudsen number is noticed when the baryonless medium changes to baryon asymmetric medium. Furthermore, the Lorenz number gets decreased due to the expansion-induced anisotropy and shows linear enhancement with the temperature, whose magnitude in the baryon asymmetric medium is smaller than that in the baryonless medium. Since its value remains above unity and it grows with the temperature, the heat transport prevails over the charge transport in an expansion-induced anisotropic medium, like in an isotropic medium. This increase 
in the Lorenz number is more evident at low temperatures, whereas at high temperatures it gets nearly saturated. 

The present work is organized as follows. In section 2, we have studied the charge transport by 
calculating the electrical conductivity for an anisotropic hot QCD medium at finite baryon 
asymmetry. In section 3, we have studied the heat transport by calculating the thermal conductivity in the similar regime. The observables, such as the Knudsen number and the Lorenz number associated with the abovementioned transport coefficients are explored in section 4. In section 5, we have discussed the quasiparticle model of partons for an anisotropic hot QCD medium at finite baryon asymmetry, wherein we have derived the anisotropic quasiparticle masses of partons. We have discussed the results in section 6. Finally, in section 7, the results of this work are concluded. 

\section{Electrical conductivity for an anisotropic hot QCD medium at finite baryon asymmetry}
The QGP matter formed in the early stages of ultrarelativistic heavy ion collisions may exhibit larger longitudinal expansion than the radial expansion, thus experiencing a local momentum anisotropy. 
If the momentum anisotropy is weak ($\xi<1$) with direction $\mathbf{n}$, the distribution function in anisotropic medium can be approximated as the isotropic one with the tail of distribution being curtailed \cite{Romatschke:PRD68'2003}. The distribution functions for quark, antiquark and gluon are thus rescaled as
\be\label{A.D.F.Q.}
&&f_f^\xi=\frac{N(\xi)}{e^{\beta\left(\sqrt{\omega_f^2+\xi(\mathbf{p}\cdot\mathbf{n})^2}-\mu_f\right)}+1} ~, \\ 
\label{A.D.F.A.}&&\bar{f}_f^\xi=\frac{N(\xi)}{e^{\beta\left(\sqrt{\omega_f^2+\xi(\mathbf{p}\cdot\mathbf{n})^2}+\mu_f\right)}+1} ~, \\ 
\label{A.D.F.G.}&&f_g^\xi=\frac{N(\xi)}{e^{\beta\sqrt{\omega_g^2+\xi(\mathbf{p}\cdot\mathbf{n})^2}}-1} 
~,\ee
respectively, where $N(\xi)=\sqrt{1+\xi}$ denotes the normalization factor \cite{Romatschke:PRD70'2004} and $\xi$ is the anisotropic parameter. After the Taylor series expansion up to $\mathcal{O}(\xi)$, $f_f^\xi$, $\bar{f}_f^\xi$ and $f_g^\xi$ take the following forms, 
\be\label{E.Q.}
&&f_f^\xi=f_f+\frac{\xi f_f}{2}-\frac{\xi\beta(\mathbf{p}\cdot\mathbf{n})^2}{2\omega_f}f_f\left(1-f_f\right), \\ 
\label{E.A.}&&\bar{f}_f^\xi=\bar{f}_f+\frac{\xi \bar{f}_f}{2}-\frac{\xi\beta(\mathbf{p}\cdot\mathbf{n})^2}{2\omega_f}\bar{f}_f\left(1-\bar{f}_f\right), \\ 
\label{E.G.}&&f_g^\xi=f_g+\frac{\xi f_g}{2}-\frac{\xi\beta(\mathbf{p}\cdot\mathbf{n})^2}{2\omega_g}f_g\left(1+f_g\right)
,\ee
where $f_f$, $\bar{f}_f$ and $f_g$ represent the isotropic quark, antiquark and gluon distribution functions, which are given by
\be\label{I.D.F.Q.}
&&f_f=\frac{1}{e^{\beta\left(\omega_f-\mu_f\right)}+1} ~, \\ 
&&\label{I.D.F.A.Q.}\bar{f}_f=\frac{1}{e^{\beta\left(\omega_f+\mu_f\right)}+1} ~, \\ 
&&\label{I.D.F.G.}f_g=\frac{1}{e^{\beta\omega_g}-1}
~.\ee
In the above equations, $\mu_f$ and $\omega_f$ denote the chemical potential and the energy of the $f$th flavor of quark (antiquark), respectively and $\omega_g$ represents the energy of the gluon in the baryon asymmetric hot QCD medium. The anisotropic parameter explains the degree of anisotropy and its general form is defined in terms of the transverse and longitudinal components of momentum as
\be\label{parameter}
\xi=\frac{\left\langle\mathbf{p}_T^2\right\rangle}{2\left\langle p_L^2\right\rangle}-1
~,\ee
where $p_L=\mathbf{p}\cdot\mathbf{n}$, $\mathbf{p}_T=\mathbf{p}-\mathbf{n}\cdot(\mathbf{p}\cdot\mathbf{n})$, $\mathbf{p}\equiv(\rm{p}\sin\theta\cos\phi,\rm{p}\sin\theta\sin\phi,\rm{p}\cos\theta)$, $\mathbf{n}=(\sin\alpha,0,\cos\alpha)$, $\alpha$ is the angle between 
z-axis and direction of anisotropy, $(\mathbf{p}\cdot\mathbf{n})^2=\rm{p}^2c(\alpha,\theta,\phi)=\rm{p}^2(\sin^2\alpha\sin^2\theta\cos^2\phi+\cos^2\alpha\cos^2\theta
+\sin(2\alpha)\sin\theta\cos\theta\cos\phi)$. For $p_T\gg p_L$, $\xi$ takes positive value, which explains the removal of particles with a large momentum component along the $\mathbf{n}$ direction due to the faster longitudinal expansion than the transverse expansion of matter. 

The weak-momentum anisotropy due to the asymptotic expansion of matter could affect the charge transport in the baryon asymmetric medium. This can be comprehended by calculating and observing the electrical conductivity in the similar regime. When the anisotropic hot medium of quarks, antiquarks and gluons with finite baryon asymmetry in the thermal equilibrium gets disturbed infinitesimally by an external electric field, an electric four-current is induced, which is given by
\begin{eqnarray}\label{current}
J^\mu = \sum_f g_f \int\frac{d^3\rm{p}}{(2\pi)^3\omega_f}p^\mu \left[q_f\delta f^\xi_f+{\bar q_f}\delta \bar{f}^\xi_f\right]
.\end{eqnarray}
Here, $g_f$, $q_f$ ($\bar{q}_f$) and $\delta f^\xi_f$ ($\delta \bar{f}^\xi_f$) represent the degeneracy factor, electric charge and infinitesimal change in the anisotropic quark (antiquark) distribution function of $f$th flavor, respectively. The electric charges of $u$, $d$ and $s$ quark flavors are given by $q_u=+\frac{2}{3}e$, $q_d=-\frac{1}{3}e$ and $q_s=-\frac{1}{3}e$. Further, the degeneracy factor for the $f$th flavor of quark (antiquark) is given by $g_f=2\times3=6$, where 2 is the spin degeneracy factor and 3 represents three colors. According to the Ohm's law, the electric current density (the longitudinal component of the spatial part of electric four-current) is directly proportional to the external electric field with the proportionality factor being the electrical conductivity. Thus, we have 
\be\label{Ohm's law}
\mathbf{J}=\sigma_{\rm el}\mathbf{E}
~.\ee
The infinitesimal changes in the anisotropic quark and antiquark distribution functions are respectively defined as $\delta f_f^\xi=f_f^{\prime \xi}-f_f^\xi$ and 
$\delta \bar{f}_f^\xi=\bar{f}_f^{\prime \xi}-\bar{f}_f^\xi$, where $f_f^\xi$ and $\bar{f}_f^\xi$ are the 
equilibrium distribution functions in the anisotropic medium for $f$th flavor of quark and antiquark, respectively. By solving the following relativistic Boltzmann transport equation (RBTE) in the relaxation time approximation, $\delta f_f^\xi$ can be obtained. 
\be\label{R.B.T.E.}
p^\mu\frac{\partial f_f^{\prime\xi}}{\partial x^\mu}+q_f F^{\rho\sigma} 
p_\sigma \frac{\partial f_f^{\prime\xi}}{\partial p^\rho}=-\frac{p_\nu u^\nu}{\tau_f}\delta f_f^\xi
~,\ee
where $F^{\rho\sigma}$ denotes the electromagnetic field strength tensor, $u^\nu$ represents the four-velocity of fluid in the local rest frame and the relaxation time for quark (antiquark) of $f$th 
flavor, $\tau_f$ ($\tau_{\bar{f}}$) is given \cite{Hosoya:NPB250'1985} by
\begin{eqnarray}
\tau_{f(\bar{f})}=\frac{1}{5.1T\alpha_s^2\log\left(1/\alpha_s\right)\left[1+0.12(2N_f+1)\right]}
~.\end{eqnarray}
In the present study, we utilize this momentum-independent relaxation time 
within the relaxation time approximation, a standard and effective 
approach used in many transport analyses of the QGP. The relaxation time used here acts as an effective phenomenological input and this choice allows for a transparent exploration of the qualitative influence of anisotropy and baryon asymmetry without invoking additional uncertainties from model-dependent cross-section inputs. While such an approximation is widely used for studying qualitative trends in the QGP transport phenomena, a complete microscopic treatment would require computing the relaxation time from scattering cross sections among quasiparticles, that would naturally depend on their medium-modified masses. The present formulation, therefore, can be regarded as a first approximation within the quasiparticle framework. For completeness, we note that, such approaches involving the mass-dependent relaxation times have been discussed in references \cite{Kadam:PRD98'2018,Moreau:PRC100'2019,Mykhaylova:PRD100'2019,Soloveva:PRC101'2020}. The components of $F^{\rho\sigma}$ are associated with the electric and magnetic fields. In the absence of magnetic field, in order to see the response of electric field, we use the components of $F^{\rho\sigma}$ associated with only electric field. Additionally, in case of a spatially homogeneous distribution function with the steady-state condition, we use $\frac{\partial f_f^{\prime\xi}}{\partial \mathbf{r}}=0$ and 
$\frac{\partial f_f^{\prime\xi}}{\partial t}=0$. Thus the RBTE \eqref{R.B.T.E.} takes the following form, 
\be\label{R.B.T.E.(1)}
q_f\mathbf{E}\cdot\mathbf{p}\frac{\partial f_f^{\prime\xi}}{\partial p_0}+q_f p_0\mathbf{E}\cdot\frac{\partial f_f^{\prime\xi}}{\partial \mathbf{p}}=-\frac{p_0}{\tau_f}\delta f_f^\xi
~.\ee
After solving the RBTE \eqref{R.B.T.E.(1)}, we get $\delta f_f^\xi$ for quark as
\be
\delta f_f^\xi=\frac{2\tau_f \beta q_f\mathbf{E}\cdot\mathbf{p}}{\omega_f}f_f\left(1-f_f\right)\left[1+\frac{\xi}{2}+\frac{\xi c(\alpha,\theta,\phi)}{2}\left(1+\frac{2\beta{\rm p}^2f_f}{\omega_f}-\frac{\beta\rm{p}^2}{\omega_f}-\frac{\rm{p}^2}{\omega_f^2}\right)\right]
.\ee
Similarly, $\delta \bar{f}_f^\xi$ for antiquark is determined as
\be
\delta \bar{f}_f^\xi=\frac{2\tau_{\bar{f}} \beta \bar{q}_f\mathbf{E}\cdot\mathbf{p}}{\omega_f}\bar{f}_f\left(1-\bar{f}_f\right)\left[1+\frac{\xi}{2}+\frac{\xi c(\alpha,\theta,\phi)}{2}\left(1+\frac{2\beta{\rm p}^2\bar{f}_f}{\omega_f}-\frac{\beta\rm{p}^2}{\omega_f}-\frac{\rm{p}^2}{\omega_f^2}\right)\right]
.\ee
Substituting $\delta f_f^\xi$ and $\delta \bar{f}_f^\xi$ in the spatial component of eq. (\ref{current}) 
and then comparing with eq. \eqref{Ohm's law}, we get the electrical conductivity for an 
expansion-induced anisotropic medium at finite baryon asymmetry as
\be\label{A.E.C.}
\nonumber\sigma_{\rm el} &=& \frac{\beta}{3\pi^2}\sum_f g_f q_f^2\int d{\rm p} ~ \frac{{\rm p}^4}{\omega_f^2}\left[\tau_f f_f\left(1-f_f\right)+\tau_{\bar{f}} \bar{f}_f\left(1-\bar{f}_f\right)\right] \\ && \nonumber+\frac{\xi\beta}{6\pi^2}\sum_f g_f q_f^2\int d{\rm p} ~ \frac{{\rm p}^4}{\omega_f^2}\left[\tau_f f_f\left(1-f_f\right)+\tau_{\bar{f}} \bar{f}_f\left(1-\bar{f}_f\right)\right] \\ && \nonumber -\frac{\xi\beta^2}{18\pi^2}\sum_f g_f q_f^2\int d{\rm p} ~ \frac{{\rm p}^6}{\omega_f^3}\left[\tau_f f_f\left(1-f_f\right)+\tau_{\bar{f}} \bar{f}_f\left(1-\bar{f}_f\right)\right] \\ && \nonumber+\frac{\xi\beta^2}{9\pi^2}\sum_f g_f q_f^2\int d{\rm p} ~ \frac{{\rm p}^6}{\omega_f^3}\left[\tau_f f^2_f\left(1-f_f\right)+\tau_{\bar{f}} \bar{f}^2_f\left(1-\bar{f}_f\right)\right] \\ && \nonumber -\frac{\xi\beta}{18\pi^2}\sum_f g_f q_f^2\int d{\rm p} ~ \frac{{\rm p}^6}{\omega_f^4}\left[\tau_f f_f\left(1-f_f\right)+\tau_{\bar{f}} \bar{f}_f\left(1-\bar{f}_f\right)\right] \\ && +\frac{\xi\beta}{18\pi^2}\sum_f g_f q_f^2\int d{\rm p} ~ \frac{{\rm p}^4}{\omega_f^2}\left[\tau_f f_f\left(1-f_f\right)+\tau_{\bar{f}} \bar{f}_f\left(1-\bar{f}_f\right)\right]
.\ee 
The first term in the right-hand side of the above equation is the electrical conductivity ($\sigma_{\rm el}^i$) for an isotropic medium ($\xi=0$). Thus, in terms of $\xi$-independent ($\sigma_{\rm el}^i$) and $\xi$-dependent ($\sigma_{\rm el}^\xi$) parts, $\sigma_{\rm el}$ is written as
\be\label{A.E.C.(1)}
\nonumber\sigma_{\rm el} &=& \sigma_{\rm el}^i+\sigma_{\rm el}^\xi \\ &=& \nonumber\sigma_{\rm el}^i-\xi\left[\frac{\beta^2}{18\pi^2}\sum_f g_f q_f^2\int d{\rm p} ~ \frac{{\rm p}^6}{\omega_f^3}\left\lbrace\tau_f f_f\left(1-f_f\right)\left(1-2f_f+\frac{1}{\beta\omega_f}\right)\right.\right. \\ && \left.\left.\nonumber+\tau_{\bar{f}} \bar{f}_f\left(1-\bar{f}_f\right)\left(1-2\bar{f}_f+\frac{1}{\beta\omega_f}\right)\right\rbrace\right. \\ && \left.-\frac{2\beta}{9\pi^2}\sum_f g_f q_f^2\int d{\rm p} ~ \frac{{\rm p}^4}{\omega_f^2}\left\lbrace\tau_f f_f\left(1-f_f\right)+\tau_{\bar{f}} \bar{f}_f\left(1-\bar{f}_f\right)\right\rbrace\right]
.\ee

\section{Thermal conductivity for an anisotropic hot QCD medium at finite baryon asymmetry}
The heat transport phenomenon in a baryon asymmetric hot QCD medium can be comprehended by the 
corresponding transport coefficient, {\em i.e.} the thermal conductivity. The heat flow 
four-vector is defined as the difference between the energy diffusion and the enthalpy diffusion as
\be\label{heat flow}
Q_\mu=\Delta_{\mu\alpha}T^{\alpha\beta}u_\beta-h\Delta_{\mu\alpha}N^\alpha
,\ee
where $T^{\alpha\beta}$ is the energy-momentum tensor, $N^\alpha$ is the particle flow four-vector, $\Delta_{\mu\alpha}=g_{\mu\alpha}-u_\mu u_\alpha$, the enthalpy per particle $h=(\varepsilon+P)/n$ 
with $\varepsilon$, $P$ and $n$ representing the energy density, the pressure and the particle number density, respectively. Further, $N^\alpha$ and $T^{\alpha\beta}$ are also known as the first and the second moments 
of the distribution function, respectively with the following expressions, 
\be
&&N^\alpha=\sum_f g_f\int \frac{d^3{\rm p}}{(2\pi)^3\omega_f}p^\alpha \left[f_f^\xi+\bar{f}_f^\xi\right] \label{P.F.F.}, \\ 
&&T^{\alpha\beta}=\sum_f g_f\int \frac{d^3{\rm p}}{(2\pi)^3\omega_f}p^\alpha p^\beta \left[f_f^\xi+\bar{f}_f^\xi\right] \label{E.M.T.}
.\ee
Using $N^\alpha$ and $T^{\alpha\beta}$, we get $n=N^\alpha u_\alpha$, $\varepsilon=u_\alpha T^{\alpha\beta} u_\beta$ and $P=-\Delta_{\alpha\beta}T^{\alpha\beta}/3$. In the local rest frame, the heat flow is purely spatial, because $Q_\mu u^\mu=0$. Hence, the spatial component of the heat flow four-vector remains finite and is given by
\be\label{heat1 (1)}
Q^i=\sum_f g_f\int \frac{d^3{\rm p}}{(2\pi)^3} ~ \frac{p^i}{\omega_f}\left[(\omega_f-h_f)\delta f_f^\xi+(\omega_f-\bar{h}_f)\delta \bar{f}_f^\xi\right]
.\ee
Through the Navier-Stokes equation, the heat flow is related to the gradients of temperature and pressure as
\be\label{heat.1}
Q^i=-\kappa\delta^{ij}\left[\partial_j T - \frac{T}{\varepsilon+P}\partial_j P\right] 
,\ee
where $\kappa$ is the thermal conductivity. For the determination of the thermal conductivity, the electromagnetic field strength part is not needed, thus, it can be dropped from the relativistic 
Boltzmann transport equation (\ref{R.B.T.E.}). Expanding the gradient of the anisotropic quark distribution function in terms of the gradients of the flow velocity and the temperature in the relaxation time approximation and simplifying, we get 
\be\label{eq1.1}
&&\nonumber \left[1+\frac{\xi}{2}-\frac{\xi\beta{\rm p}^2c(\alpha,\theta,\phi)}{2p_0}\left(1-2f_f\right)\right]p^\mu\partial_\mu f_f+\frac{\xi\beta{\rm p}^2c(\alpha,\theta,\phi)}{2p_0}f_f\left(1-f_f\right)\left(p^\alpha Du_\alpha\right. \\ && \left.+\frac{p^\mu p^\alpha}{p_0}\nabla_\mu u_\alpha+\beta p_0DT+\beta p^\mu\nabla_\mu T\right) = -\frac{p_\nu u^\nu}{\tau_f}\delta f_f^\xi
.\ee
After solving eq. (\ref{eq1.1}), $\delta f_f^\xi$ is obtained as
\be\label{delta.q1}
\nonumber\delta f_f^\xi &=& -\beta\tau_ff_f\left(1-f_f\right)\left[1+\frac{\xi}{2}-\frac{\xi\beta{\rm p}^2c(\alpha,\theta,\phi)}{2p_0}\left(1-2f_f\right)\right]\left[p_0\frac{DT}{T}+TD\left(\frac{\mu}{T}\right)\right. \\ && \left.\nonumber+\frac{\left(\omega_f-h_f\right)p^j}{T\omega_f}\left(\partial_jT-\frac{T}{\varepsilon+P}\partial_jP\right)-\frac{p^\mu p^\alpha}{p_0}\nabla_\mu u_\alpha\right]-\frac{\xi\beta\tau_f{\rm p}^2c(\alpha,\theta,\phi)}{2p^2_0} \\ && \times f_f\left(1-f_f\right)\left(p^\alpha Du_\alpha+\frac{p^\mu p^\alpha}{p_0}\nabla_\mu u_\alpha+\beta p_0DT+\beta p^\mu\nabla_\mu T\right)
.\ee
Similarly, $\delta \bar{f}_f^\xi$ for antiquark is determined as
\be\label{delta.aq1}
\nonumber\delta \bar{f}_f^\xi &=& -\beta\tau_{\bar{f}}\bar{f}_f\left(1-\bar{f}_f\right)\left[1+\frac{\xi}{2}-\frac{\xi\beta{\rm p}^2c(\alpha,\theta,\phi)}{2p_0}\left(1-2\bar{f}_f\right)\right]\left[p_0\frac{DT}{T}-TD\left(\frac{\mu}{T}\right)\right. \\ && \left.\nonumber+\frac{\left(\omega_f-\bar{h}_f\right)p^j}{T\omega_f}\left(\partial_jT-\frac{T}{\varepsilon+P}\partial_jP\right)-\frac{p^\mu p^\alpha}{p_0}\nabla_\mu u_\alpha\right]-\frac{\xi\beta\tau_{\bar{f}}{\rm p}^2c(\alpha,\theta,\phi)}{2p^2_0} \\ && \times\bar{f}_f\left(1-\bar{f}_f\right)\left(p^\alpha Du_\alpha+\frac{p^\mu p^\alpha}{p_0}\nabla_\mu u_\alpha+\beta p_0DT+\beta p^\mu\nabla_\mu T\right)
.\ee
Substituting the values of $\delta f_f^\xi$ and $\delta \bar{f}_f^\xi$ in eq. (\ref{heat1 (1)}) and then comparing with eq. (\ref{heat.1}), we get the thermal conductivity for an expansion-induced anisotropic 
medium at finite baryon asymmetry as
\be\label{A.T.C.}
\nonumber\kappa &=& \frac{\beta^2}{6\pi^2}\sum_f g_f\int d{\rm p} ~ \frac{{\rm p}^4}{\omega_f^2}\left[\tau_f\left(\omega_f-h_f\right)^2f_f\left(1-f_f\right)+\tau_{\bar{f}}\left(\omega_f-\bar{h}_f\right)^2\bar{f}_f\left(1-\bar{f}_f\right)\right] \\ && \nonumber+\frac{\xi\beta^2}{12\pi^2}\sum_fg_f\int d{\rm p} ~ \frac{{\rm p}^4}{\omega_f^2}\left[\tau_f\left(\omega_f-h_f\right)^2f_f\left(1-f_f\right)+\tau_{\bar{f}}\left(\omega_f-\bar{h}_f\right)^2\bar{f}_f\left(1-\bar{f}_f\right)\right] \\ && - \nonumber\frac{\xi\beta^3}{36\pi^2}\sum_fg_f\int d{\rm p} ~ \frac{{\rm p}^6}{\omega_f^3}\left[\tau_f\left(\omega_f-h_f\right)^2f_f\left(1-f_f\right)\left(1-2f_f\right)\right. \\ && \left.+\tau_{\bar{f}}\left(\omega_f-\bar{h}_f\right)^2\bar{f}_f\left(1-\bar{f}_f\right)\left(1-2\bar{f}_f\right)\right]
.\ee
The first term in the right-hand side of the above equation is the thermal conductivity ($\kappa^i$) for an isotropic medium ($\xi=0$). Thus, in terms of $\xi$-independent ($\kappa^i$) and $\xi$-dependent ($\kappa^\xi$) parts, $\kappa$ is written as
\be\label{A.T.C.(1)}
\nonumber\kappa &=& \kappa^i+\kappa^\xi \\ &=& \nonumber\kappa^i+\xi\left[\frac{\beta^2}{12\pi^2}\sum_fg_f\int d{\rm p} ~ \frac{{\rm p}^4}{\omega_f^2}\left\lbrace\tau_f\left(\omega_f-h_f\right)^2f_f\left(1-f_f\right)+\tau_{\bar{f}}\left(\omega_f-\bar{h}_f\right)^2\bar{f}_f\left(1-\bar{f}_f\right)\right\rbrace\right. \\ && \left. \nonumber -\frac{\beta^3}{36\pi^2}\sum_fg_f\int d{\rm p} ~ \frac{{\rm p}^6}{\omega_f^3}\left\lbrace\tau_f\left(\omega_f-h_f\right)^2f_f\left(1-f_f\right)\left(1-2f_f\right)\right.\right. \\ && \left.\left.+\tau_{\bar{f}}\left(\omega_f-\bar{h}_f\right)^2\bar{f}_f\left(1-\bar{f}_f\right)\left(1-2\bar{f}_f\right)\right\rbrace\right]
.\ee

\section{Observables}
In this section, we have studied how the observables related to the charge transport and the heat transport phenomena in a baryon asymmetric hot QCD medium get modulated by the expansion-induced anisotropy. In particular, we have explored the Knudsen number delineating the validity of local equilibrium property of the medium and the Lorenz number explicating the correlation between the heat flow and the charge flow through the Wiedemann-Franz law in subsections 4.1 and 4.2, respectively. 

\subsection{Knudsen number}
Through the Knudsen number ($\Omega$), the degree of separation between the microscopic and macroscopic length scales of the medium can be quantified. The Knudsen number is defined as the ratio of the mean free path ($\lambda$) to the characteristic length ($l$) of the medium, {\em i.e.} $\Omega={\lambda}/{l}$. For the applicability of equilibrium hydrodynamics, these two length scales must be sufficiently separated, 
{\em i.e.} $\Omega$ requires to be small or less than unity. Thus, for an equilibrium system, the mean free path should be smaller than the characteristic length of the medium. As the mean free path is 
related to the thermal conductivity through the relation $\lambda={3\kappa}/{(vC_V)}$, the Knudsen number is rewritten as
\be\label{Knudsen number}
\Omega=\frac{3\kappa}{lvC_V}
~,\ee
where $v$ and $C_V$ represent the relative speed and the specific heat at constant volume, respectively. 
In this work, we have used $v\simeq 1$ and $l=4$ fm, and determined $C_V$ from the 
energy-momentum tensor through the relation, $C_V=\partial (u_\mu T^{\mu\nu}u_\nu)/\partial T$. It would be interesting to see how the introduction of expansion-induced anisotropy affects the equilibrium 
property through the Knudsen number of the medium. 

\subsection{Lorenz number}
Through the Wiedemann-Franz law, the Lorenz number elucidates the correlation between the heat flow and the charge flow in a medium. The Wiedemann-Franz law states that the ratio of charged particle contribution 
of the thermal conductivity to the electrical conductivity is proportional to the temperature, {\em i.e.}, 
\be\label{Lorenz number}
\frac{\kappa}{\sigma_{\rm el}}=LT
~,\ee
where the proportionality factor $L$ represents the Lorenz number. This law is perfectly satisfied by the substances or the media which are good thermal and electrical conductors, such as metals. In this work, 
we intend to explore how the Lorenz number for a baryon asymmetric hot QCD medium gets modified by the expansion-induced anisotropy. It would be entrancing to understand how the emergence of aforesaid anisotropy alters the competition between the thermal and electrical conductivities through the Lorenz number of the medium. 

\section{Quasiparticle description of a baryon asymmetric hot QCD medium in the presence of expansion-induced anisotropy}
In the quasiparticle description of QGP, each constituent acquires the medium generated mass due to its interaction with other constituents of the medium, thus elucidating the collective properties of the 
medium. The thermal mass or the quasiparticle mass of a parton can get significantly altered by the presence of different extreme conditions and anisotropies. In the semiclassical transport theory, the thermal masses (squared) of quark and gluon for a baryon asymmetric hot QCD medium are respectively defined \cite{Kelly:PRD50'1994,Litim:PR364'2002} as
\be\label{Q.P.M.Q.(definition of quark mass)}
\nonumber m_{fT}^2 &=& \frac{g^2\left(N_c^2-1\right)}{2N_c}\int\frac{d^3{\rm p}}{(2\pi)^3} ~ \frac{1}{\rm p}\left[f_g+\frac{1}{2}\left(f_f+\bar{f}_f\right)\right] \\ &=& \frac{g^2\left(N_c^2-1\right)}{8\pi^2N_c}\int d{\rm p} ~ {\rm p}\left[\frac{\rm 2}{\left(e^{\beta\omega_g}-1\right)}+\frac{1}{\left(e^{\beta(\omega_f-\mu_f)}+1\right)}+\frac{1}{\left(e^{\beta(\omega_f+\mu_f)}+1\right)}\right], \\ 
\label{Q.P.M.G.(definition of gluon mass)}\nonumber m_{gT}^2 &=& \frac{g^2}{2}\left[-2N_c\int\frac{d^3{\rm p}}{(2\pi)^3} ~ \frac{\partial f_g}{\partial{\rm p}}-N_f\int\frac{d^3{\rm p}}{(2\pi)^3} ~ \left(\frac{\partial f_f}{\partial{\rm p}}+\frac{\partial \bar{f}_f}{\partial{\rm p}}\right)\right] \\ &=& \nonumber\frac{g^2N_c}{2\pi^2T}\int d{\rm p} ~ \frac{{\rm p}^3}{\omega_g}\frac{e^{\beta\omega_g}}{\left(e^{\beta\omega_g}-1\right)^2}+\frac{g^2N_f}{4\pi^2T}\int d{\rm p}~\frac{{\rm p}^3}{\omega_f}\left[\frac{e^{\beta(\omega_f-\mu_f)}}{\left(e^{\beta(\omega_f-\mu_f)}+1\right)^2}\right. \\ && \left.+\frac{e^{\beta(\omega_f+\mu_f)}}{\left(e^{\beta(\omega_f+\mu_f)}+1\right)^2}\right]
.\ee
Here, $N_f=3$ and $N_c=3$ represent the number of flavors and the number of colors, respectively. Further, the bare masses of $u$, $d$ and $s$ quark flavors used in this work are 3 MeV, 5 MeV and 100 MeV, 
respectively. The integrals in the above equations can be determined by using the hard thermal loop approximation and the simplified forms of quasiparticle masses (squared) of quark and gluon up to one-loop are given \cite{Braaten:PRD45'1992,Peshier:PRD66'2002,Blaizot:PRD72'2005} by
\be\label{Q.P.M.(Quark mass)}
&& m_{fT}^2=\frac{g^2T^2}{6}\left(1+\frac{\mu_f^2}{\pi^2T^2}\right), \\
&&\label{Q.P.M.(Gluon mass)}m_{gT}^2=\frac{g^2T^2}{6}\left(N_c+\frac{N_f}{2}+\frac{3}{2\pi^2T^2}\sum_f\mu_f^2\right)
,\ee
respectively. Here, the quasiparticle masses depend on both temperature and chemical potential. The corresponding dispersion relations for quark and gluon within the quasiparticle description of an 
isotropic hot QCD medium with finite baryon asymmetry are written as
\be
&&\label{I.D.R.(Quark)}\omega_f=\sqrt{{\rm p}^2+m_{fT}^2} ~, \\ 
&&\label{I.D.R.(Gluon)}\omega_g=\sqrt{{\rm p}^2+m_{gT}^2}
~.\ee
In this regime, the quark and gluon distribution functions involve the abovementioned dispersion relations. 
We note that the chemical potentials for $u$, $d$ and $s$ quark flavors are taken to be the same in this 
work, {\em i.e.} $\mu_f=\mu$. In the above equations, $g$ denotes the one-loop running coupling with the following \cite{Kapusta:BOOK'2006} form, 
\begin{eqnarray}
g^2=4\pi\alpha_s=\frac{48\pi^2}{\left(11N_c-2N_f\right)\ln\left({\Lambda^2}/{\Lambda_{\rm\overline{MS}}^2}\right)}
~,\end{eqnarray}
where $\Lambda_{\rm\overline{MS}}=0.176$ GeV, $\Lambda=2\pi\sqrt{T^2+\mu_f^2/\pi^2}$ for electrically charged particles (quarks and antiquarks) and $\Lambda=2 \pi T$ for gluons. The presence of expansion-induced anisotropy modifies the quasiparticle masses of partons. Thus, in the presence of expansion-induced anisotropy, equations \eqref{Q.P.M.Q.(definition of quark mass)} and \eqref{Q.P.M.G.(definition of gluon mass)} get respectively modified as
\be\label{Q.P.M.Q.(Anisotropy)}
m_{fT\xi}^2 &=& \frac{g^2\left(N_c^2-1\right)}{2N_c}\int\frac{d^3{\rm p}}{(2\pi)^3} ~ \frac{1}{\rm p}\left[f_g^\xi+\frac{1}{2}\left(f_f^\xi+\bar{f}_f^\xi\right)\right], \\ 
\label{Q.P.M.G.(Anisotropy)}
m_{gT\xi}^2 &=& \frac{g^2}{2}\left[-2N_c\int\frac{d^3{\rm p}}{(2\pi)^3} ~ \frac{\partial f_g^\xi}{\partial{\rm p}}-N_f\int\frac{d^3{\rm p}}{(2\pi)^3} ~ \left(\frac{\partial f_f^\xi}{\partial{\rm p}}+\frac{\partial \bar{f}_f^\xi}{\partial{\rm p}}\right)\right]
.\ee
Substituting the values of the anisotropic distribution functions ($f_f^\xi$, $\bar{f}_f^\xi$ and $f_g^\xi$) 
in the above equations and then calculating, we obtain the quasiparticle masses (squared) of quark and gluon for an expansion-induced anisotropic hot QCD medium with finite baryon asymmetry as
\be\label{Q.P.M.Q.(Anisotropy 1)}
\nonumber m_{fT\xi}^2 &=& \frac{2g^2}{3\pi^2}\int d{\rm p} ~ {\rm p} \left[f_g+\frac{1}{2}\left(f_f+\bar{f}_f\right)\right]+\frac{\xi g^2}{3\pi^2}\int d{\rm p} ~ {\rm p} \left[f_g+\frac{1}{2}\left(f_f+\bar{f}_f\right)\right] \\ && -\frac{\xi\beta g^2}{9\pi^2}\int d{\rm p} ~ {\rm p^2} ~ f_g\left(1+f_g\right)-\frac{\xi\beta g^2}{18\pi^2}\int d{\rm p} ~ \frac{\rm p^3}{\omega_f}\left[f_f\left(1-f_f\right)+\bar{f}_f\left(1-\bar{f}_f\right)\right], \\
\label{Q.P.M.G.(Anisotropy 1)}
\nonumber m_{gT\xi}^2 &=& \frac{3\beta g^2}{2\pi^2}\left[\int d{\rm p} ~ {\rm p^2} ~ f_g\left(1+f_g\right)+\sum_f\int d{\rm p} ~ \frac{\rm p^3}{6\omega_f}\left\lbrace f_f\left(1-f_f\right)+\bar{f}_f\left(1-\bar{f}_f\right)\right\rbrace\right] \\ && \nonumber+\frac{3\xi\beta g^2}{4\pi^2}\int d{\rm p} ~ {\rm p^2} ~ f_g\left(1+f_g\right)+\frac{\xi\beta g^2}{4\pi^2}\int d{\rm p} ~ {\rm p^2} ~ f_g\left(1+f_g\right) \\ && \nonumber -\frac{\xi\beta^2 g^2}{4\pi^2}\int d{\rm p} ~ {\rm p^3}\left(1+2f_g\right)f_g\left(1+f_g\right) \\ && +\nonumber\sum_f\frac{\xi\beta g^2}{8\pi^2}\int d{\rm p} ~ \frac{\rm p^3}{\omega_f}\left[f_f\left(1-f_f\right)+\bar{f}_f\left(1-\bar{f}_f\right)\right] \\ && \nonumber+\sum_f\frac{\xi\beta g^2}{12\pi^2}\int d{\rm p} ~ \frac{\rm p^3}{\omega_f}\left[f_f\left(1-f_f\right)+\bar{f}_f\left(1-\bar{f}_f\right)\right] \\ && - \nonumber\sum_f\frac{\xi\beta g^2}{24\pi^2}\int d{\rm p} ~ \frac{\rm p^5}{\omega_f^3}\left[f_f\left(1-f_f\right)+\bar{f}_f\left(1-\bar{f}_f\right)\right] \\ && -\sum_f\frac{\xi\beta^2 g^2}{24\pi^2}\int d{\rm p} ~ \frac{\rm p^5}{\omega_f^2}\left[\left(1-2f_f\right)f_f\left(1-f_f\right)+\left(1-2\bar{f}_f\right)\bar{f}_f\left(1-\bar{f}_f\right)\right]
,\ee
respectively. We note that the quark flavors are not treated as degenerate in this work, as they differ by both their electric charges and bare masses even though their chemical potentials are taken to be identical. Further, the quasiparticle masses of different quark flavors depend on their respective bare masses through their dispersion relations ($\omega_f=\sqrt{{\rm p}^2+m^2_f}$) and the corresponding particle/antiparticle distribution functions ($f_f$ and $\bar{f}_f$). In equations \eqref{Q.P.M.Q.(Anisotropy 1)} and \eqref{Q.P.M.G.(Anisotropy 1)}, the first terms on the right-hand side are the quasiparticle masses (squared) of quark and gluon, respectively for an isotropic hot QCD medium with finite baryon asymmetry. Thus, we have 
\be\label{Q.P.M.Q.(Anisotropic medium)}
\nonumber m_{fT\xi}^2 &=& m_{fT}^2+\frac{\xi g^2}{3\pi^2}\int d{\rm p} ~ {\rm p} \left[f_g+\frac{1}{2}\left(f_f+\bar{f}_f\right)\right]-\frac{\xi\beta g^2}{9\pi^2}\int d{\rm p} ~ {\rm p^2} ~ f_g\left(1+f_g\right) \\ && -\frac{\xi\beta g^2}{18\pi^2}\int d{\rm p} ~ \frac{\rm p^3}{\omega_f}\left[f_f\left(1-f_f\right)+\bar{f}_f\left(1-\bar{f}_f\right)\right], \\
\label{Q.P.M.G.(Anisotropic medium)}
\nonumber m_{gT\xi}^2 &=& m_{gT}^2+\frac{3\xi\beta g^2}{4\pi^2}\int d{\rm p} ~ {\rm p^2} ~ f_g\left(1+f_g\right)+\frac{\xi\beta g^2}{4\pi^2}\int d{\rm p} ~ {\rm p^2} ~ f_g\left(1+f_g\right) \\ && \nonumber -\frac{\xi\beta^2 g^2}{4\pi^2}\int d{\rm p} ~ {\rm p^3}\left(1+2f_g\right)f_g\left(1+f_g\right) \\ && +\nonumber\sum_f\frac{\xi\beta g^2}{8\pi^2}\int d{\rm p} ~ \frac{\rm p^3}{\omega_f}\left[f_f\left(1-f_f\right)+\bar{f}_f\left(1-\bar{f}_f\right)\right] \\ && \nonumber+\sum_f\frac{\xi\beta g^2}{12\pi^2}\int d{\rm p} ~ \frac{\rm p^3}{\omega_f}\left[f_f\left(1-f_f\right)+\bar{f}_f\left(1-\bar{f}_f\right)\right] \\ && - \nonumber\sum_f\frac{\xi\beta g^2}{24\pi^2}\int d{\rm p} ~ \frac{\rm p^5}{\omega_f^3}\left[f_f\left(1-f_f\right)+\bar{f}_f\left(1-\bar{f}_f\right)\right] \\ && -\sum_f\frac{\xi\beta^2 g^2}{24\pi^2}\int d{\rm p} ~ \frac{\rm p^5}{\omega_f^2}\left[\left(1-2f_f\right)f_f\left(1-f_f\right)+\left(1-2\bar{f}_f\right)\bar{f}_f\left(1-\bar{f}_f\right)\right]
.\ee
It is evident from the above equations that, in an expansion-induced anisotropic medium with finite baryon asymmetry, the quasiparticle masses of partons depend on the anisotropic parameter, in addition to their dependence on the temperature and the chemical potential. Accordingly, the dispersion relations for quark and gluon get modified and take the following forms within the quasiparticle description of an anisotropic hot QCD medium with finite baryon asymmetry, 
\be
&&\label{A.D.R.(Quark)}\omega_f=\sqrt{{\rm p}^2+m_{fT\xi}^2} ~, \\ 
&&\label{A.D.R.(Gluon)}\omega_g=\sqrt{{\rm p}^2+m_{gT\xi}^2}
~.\ee
In this regime, the quark and gluon distribution functions incorporate the abovementioned dispersion relations. 

\begin{figure}[]
\begin{center}
\begin{tabular}{c c}
\includegraphics[width=7.4cm]{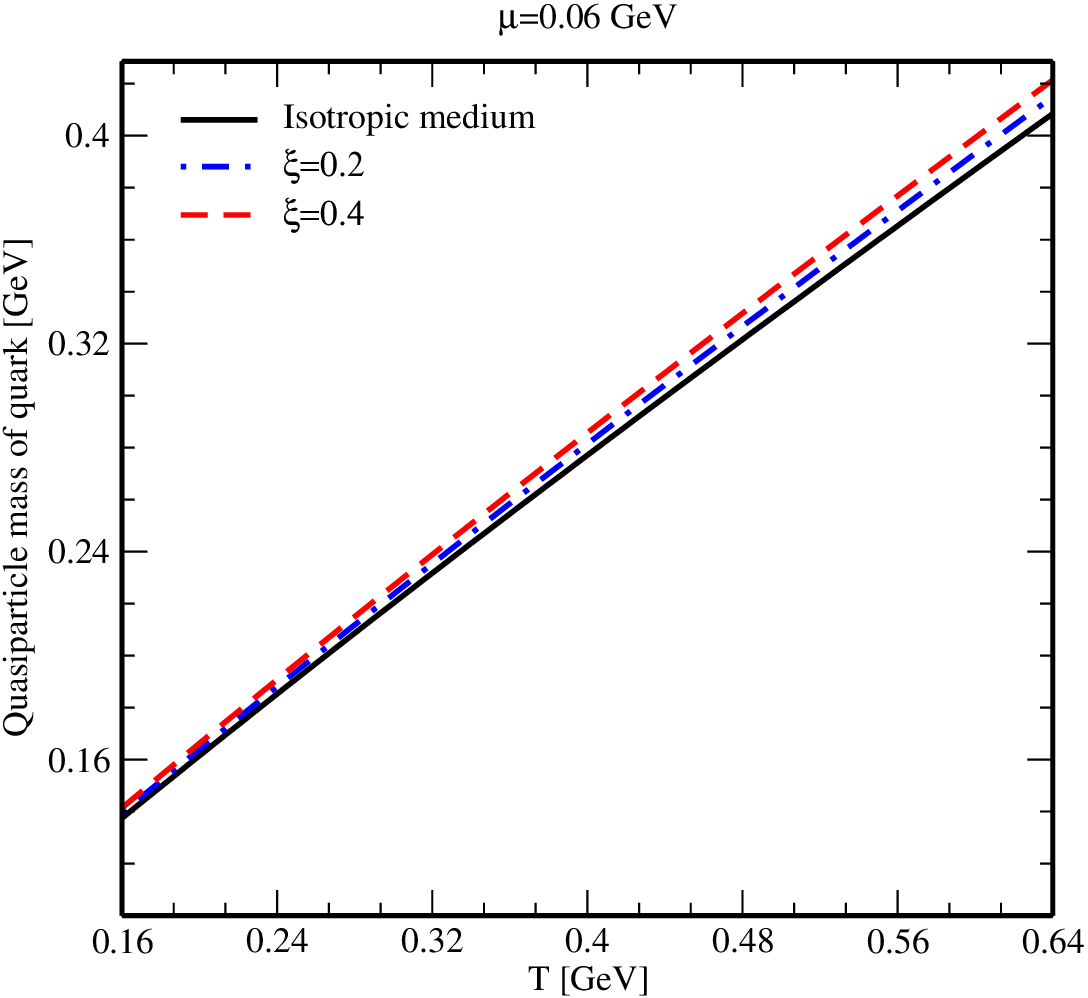}&
\hspace{0.73 cm}
\includegraphics[width=7.4cm]{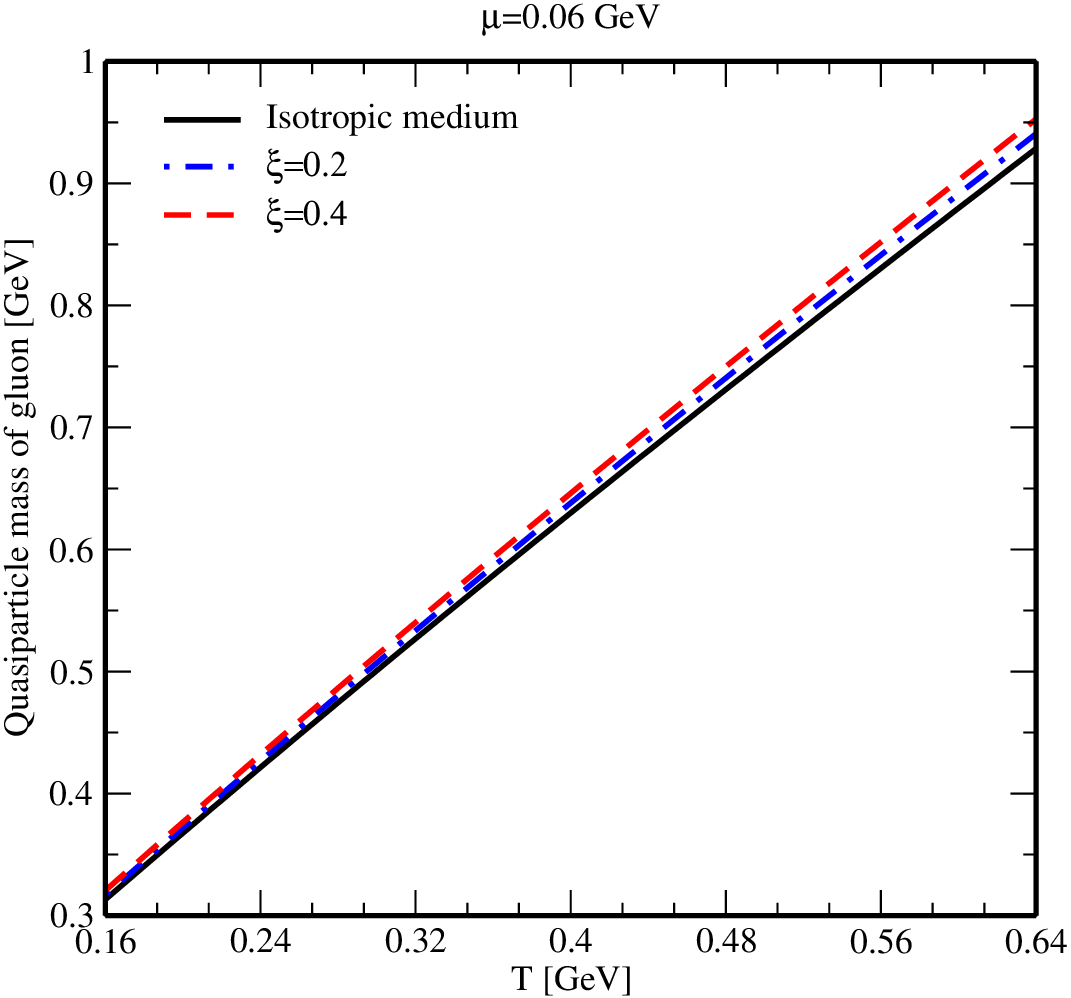} \\
a & b
\end{tabular}
\caption{Variations of (a) the quasiparticle mass of quark and (b) the quasiparticle mass of gluon 
with temperature for different values of anisotropic parameter.}\label{Fig.tmdf}
\end{center}
\end{figure}

\begin{figure}[]
\begin{center}
\begin{tabular}{c c}
\includegraphics[width=7.4cm]{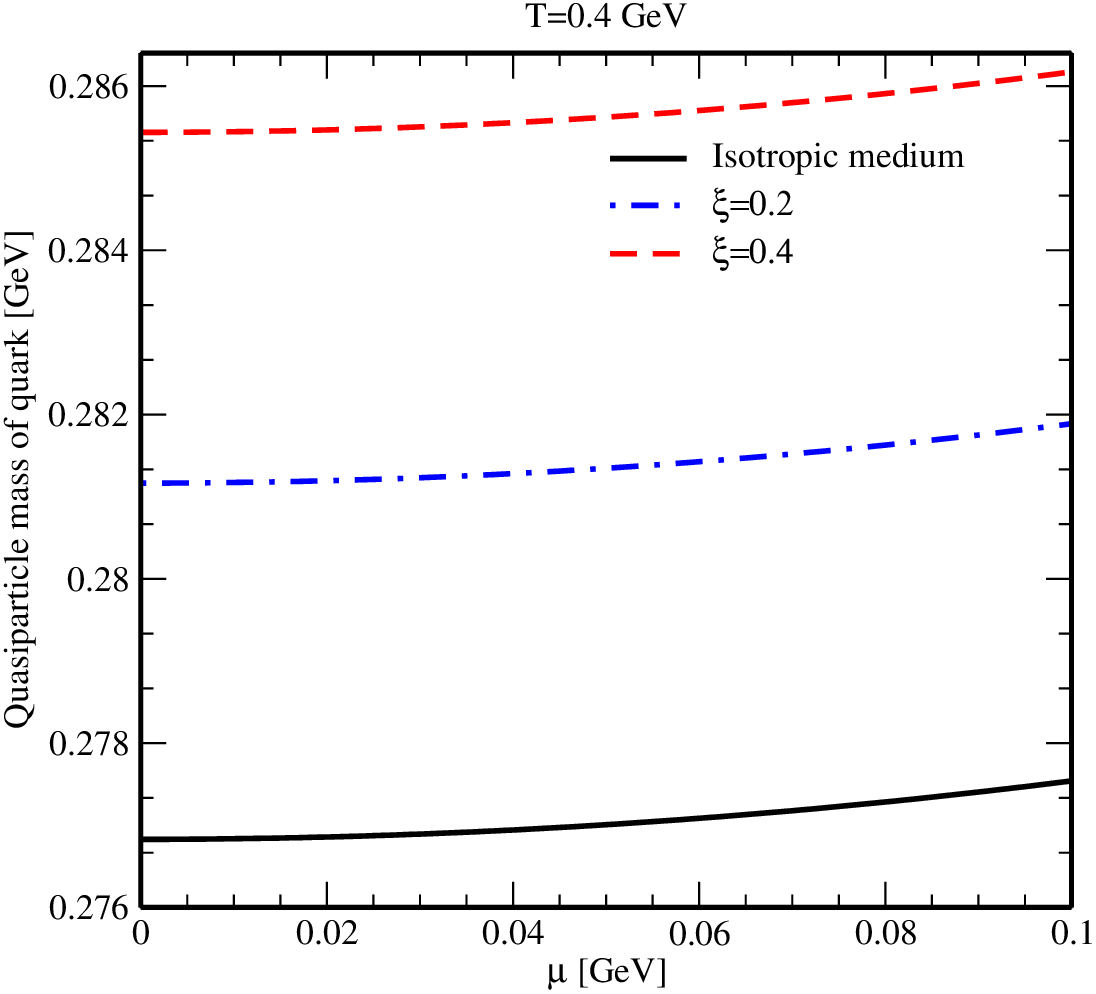}&
\hspace{0.73 cm}
\includegraphics[width=7.4cm]{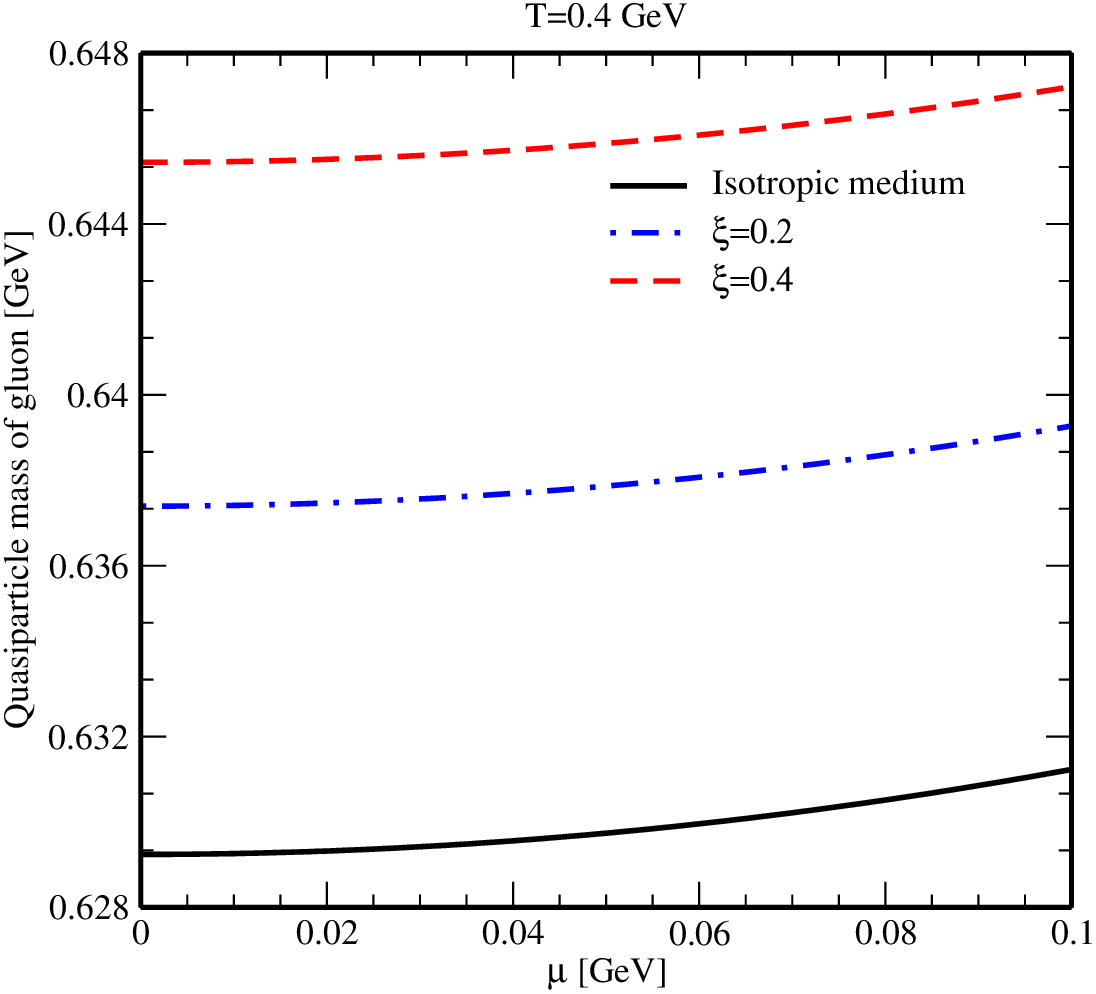} \\
a & b
\end{tabular}
\caption{Variations of (a) the quasiparticle mass of quark and (b) the quasiparticle mass of gluon 
with chemical potential for different values of anisotropic parameter.}\label{Fig.cpmdf}
\end{center}
\end{figure}

\begin{figure}[]
\begin{center}
\begin{tabular}{c c}
\includegraphics[width=7.4cm]{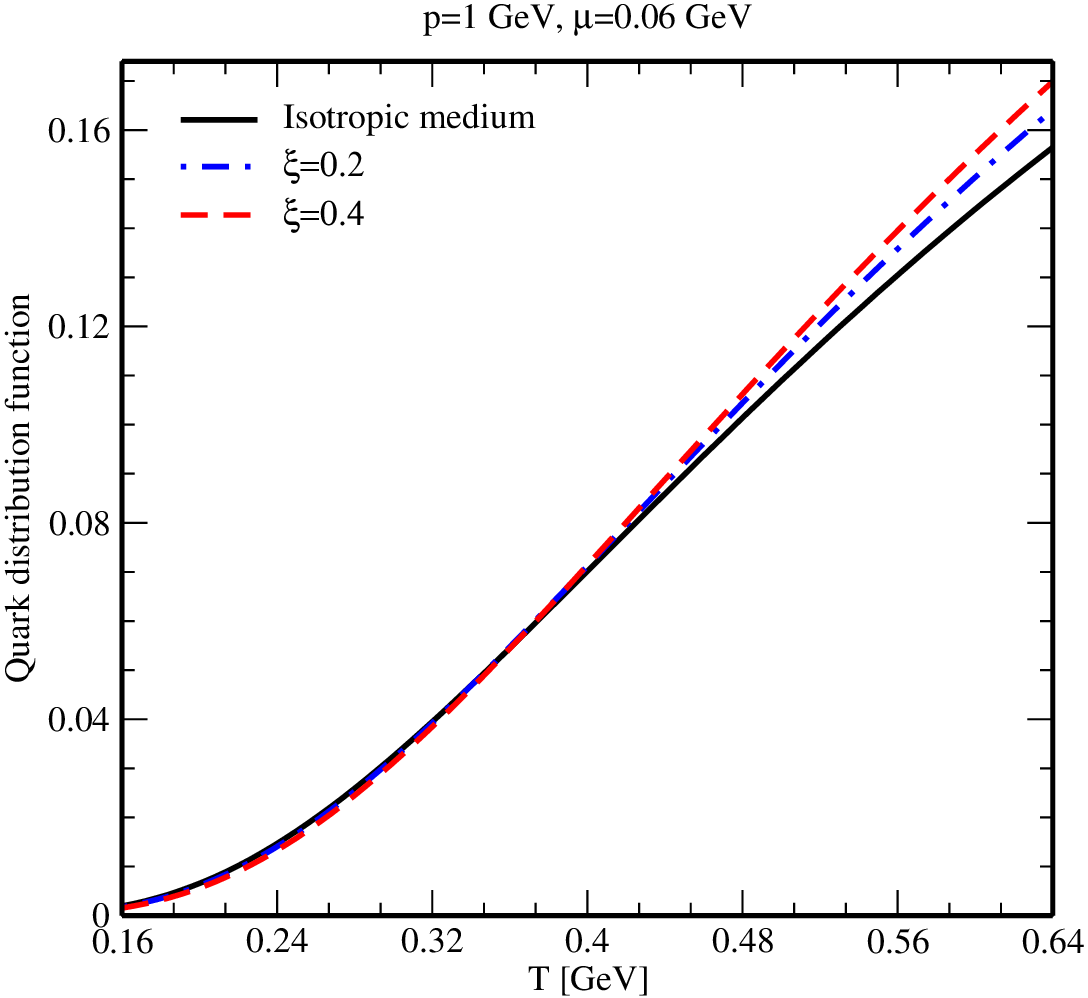}&
\hspace{0.73 cm}
\includegraphics[width=7.4cm]{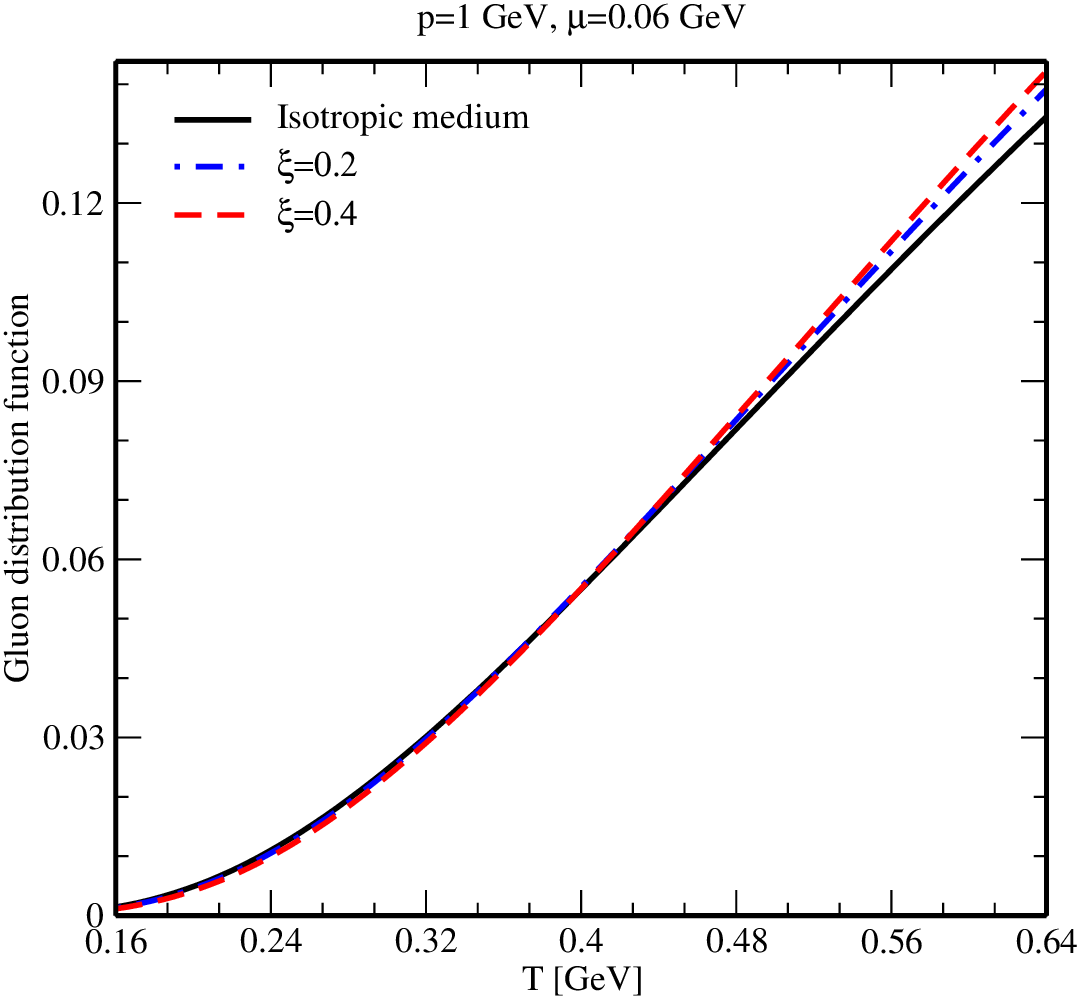} \\
a & b
\end{tabular}
\caption{Temperature dependence of (a) quark and (b) gluon distribution functions for different values of anisotropic parameter.}\label{Fig.tdf}
\end{center}
\end{figure}

\begin{figure}[]
\begin{center}
\begin{tabular}{c c}
\includegraphics[width=7.4cm]{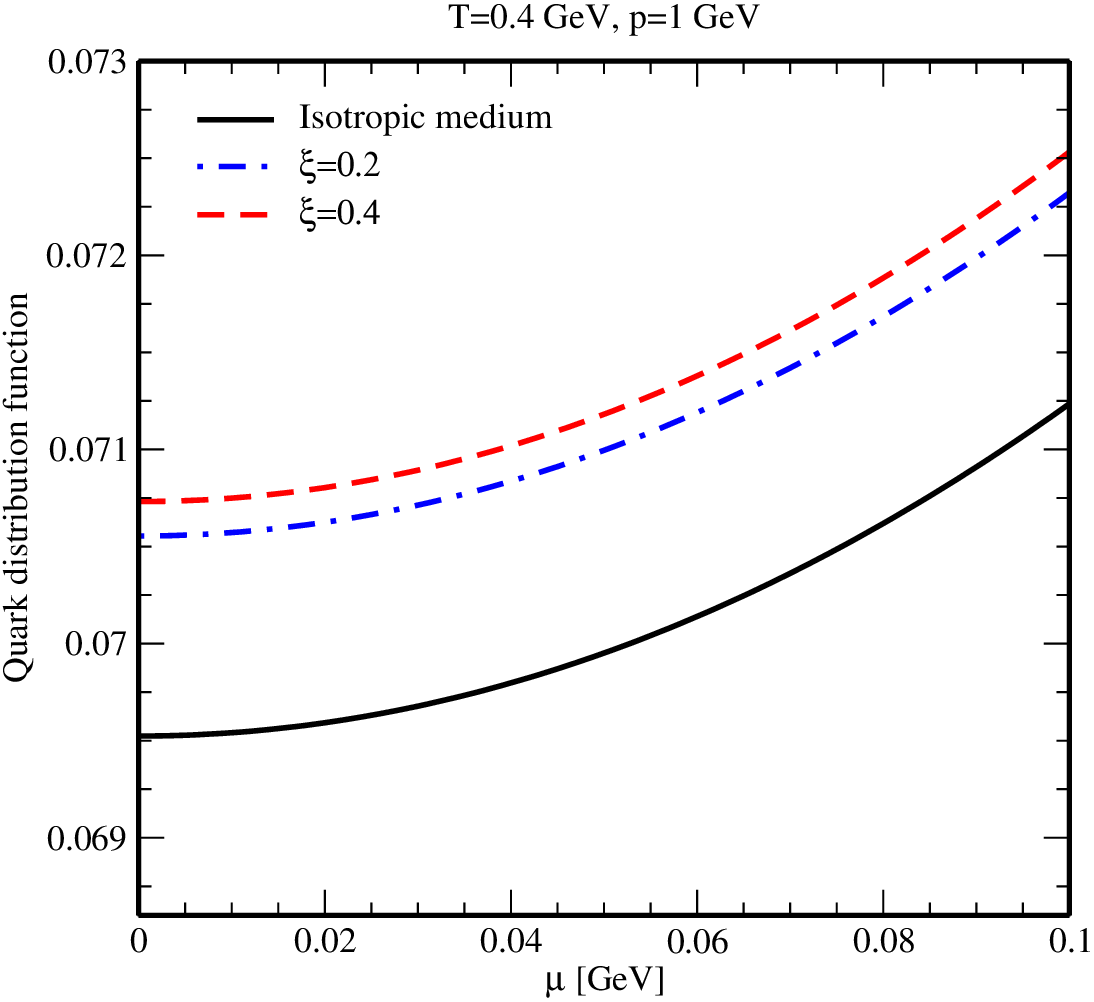}&
\hspace{0.73 cm}
\includegraphics[width=7.4cm]{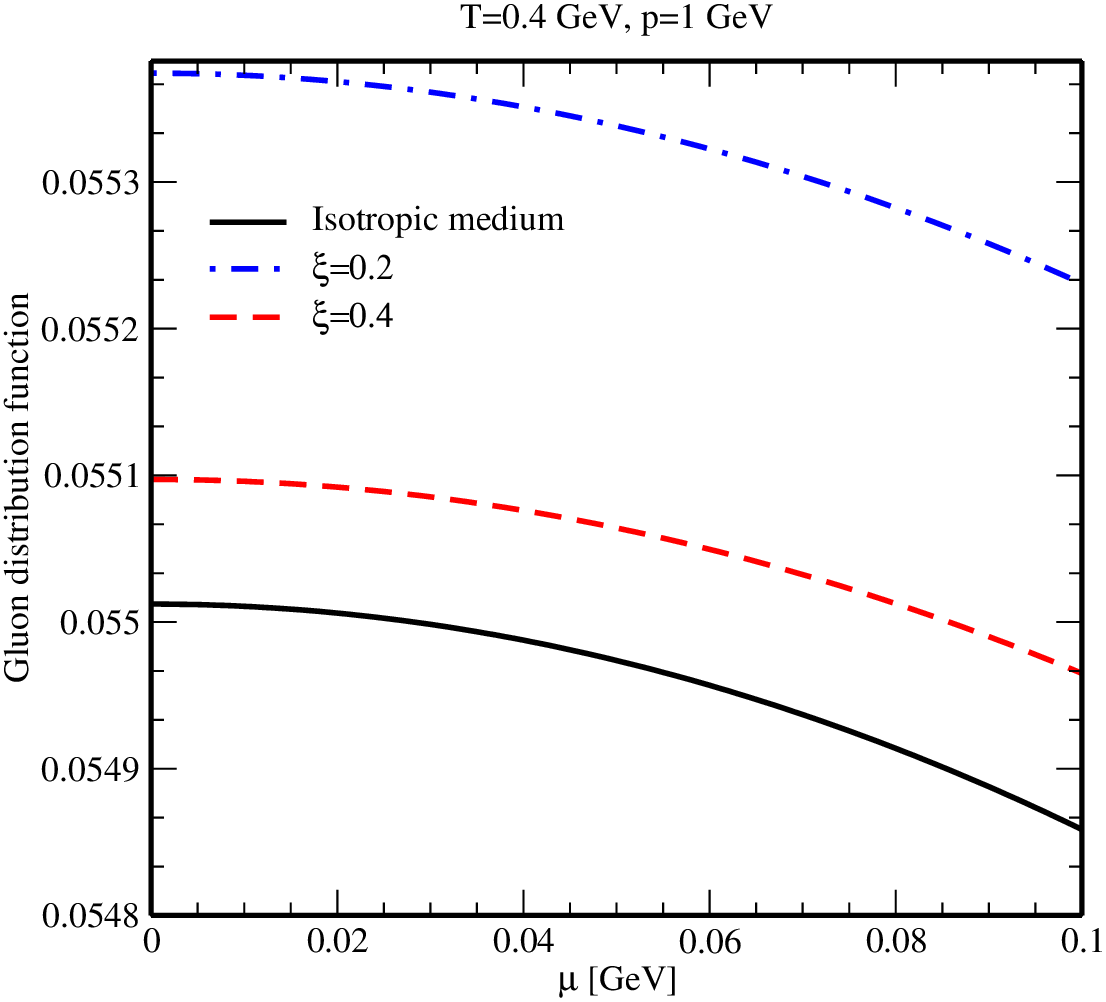} \\
a & b
\end{tabular}
\caption{Chemical potential dependence of (a) quark and (b) gluon distribution functions for different values of anisotropic parameter.}\label{Fig.cpdf}
\end{center}
\end{figure}

\begin{figure}[]
\begin{center}
\begin{tabular}{c c}
\includegraphics[width=7.4cm]{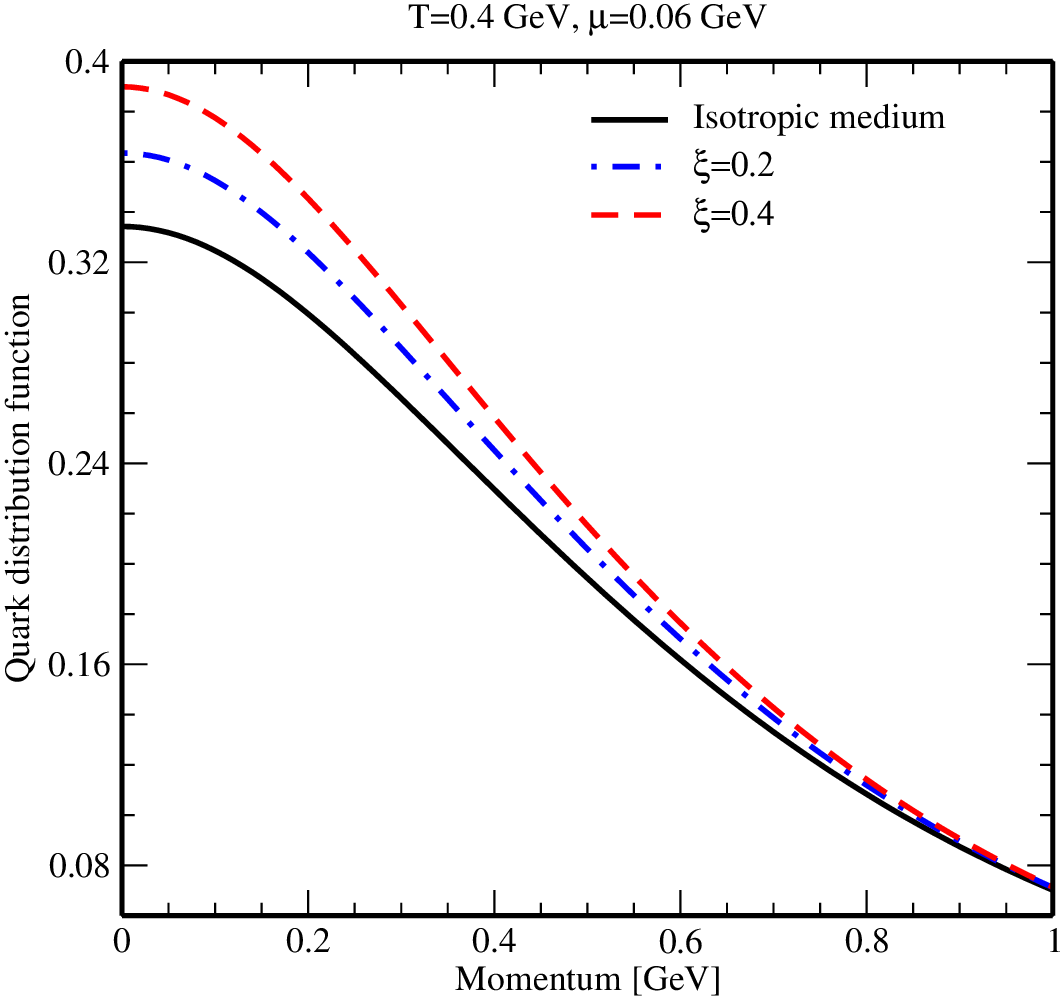}&
\hspace{0.73 cm}
\includegraphics[width=7.4cm]{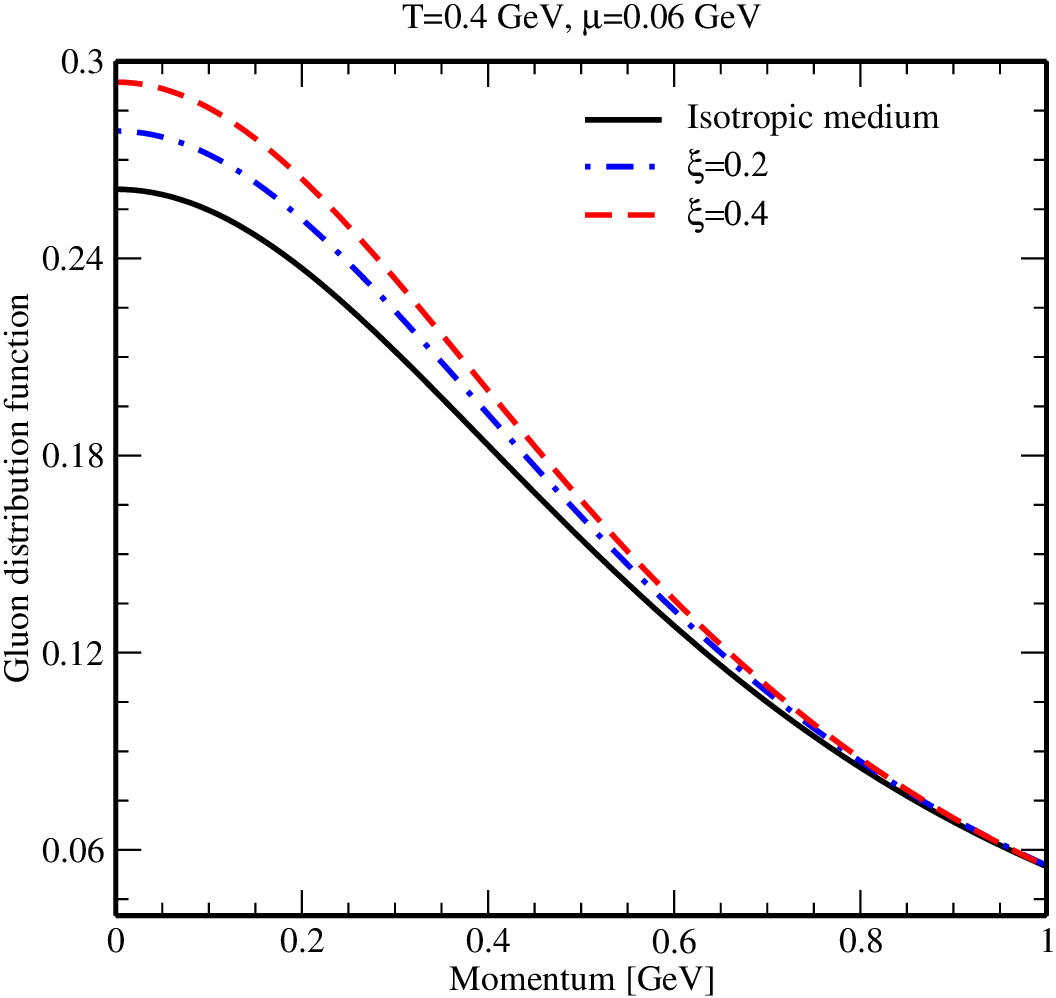} \\
a & b
\end{tabular}
\caption{Momentum dependence of (a) quark and (b) gluon distribution functions for different values of anisotropic parameter.}\label{Fig.pdf}
\end{center}
\end{figure}

Before analysing the results on the electrical conductivity, the thermal conductivity and the associated observables in the presence of anisotropy arising due to the asymptotic expansion of matter, it is 
essential to understand how the distribution functions of partons get influenced by the aforesaid anisotropy, because, within the kinetic theory framework, the transport coefficients and observables are predominantly affected by the distribution functions and the dispersion relations of partons embodying the essence of anisotropy. Therefore, it is crucial to understand how the parton distribution functions depend 
on the temperature, chemical potential and momentum in the presence of anisotropy. 

Let us start with the variations of quasiparticle masses of quark and gluon with the temperature and with the chemical potential. Figures \ref{Fig.tmdf} and \ref{Fig.cpmdf} show that the emergence of anisotropy 
enhances the quasiparticle masses of partons, thus making them more massive than their isotropic 
counterparts. This enhancement in the quasiparticle masses is more pronounced at high temperatures for a baryon asymmetric matter than that at low temperatures for a baryonless matter. Further, the quasiparticle masses exhibit a monotonic increase with the temperature when plotted at a fixed chemical potential, unlike their slow increase with the chemical potential at a fixed temperature. This temperature dependence of the quasiparticle masses over the considered temperature range can be comprehended as follows. In the high temperature regime of the QGP ($T=0.16$ GeV - $0.64$ GeV) with small chemical potential ($\mu=0.06$ GeV), the energy scale associated with the temperature dominates over other energy scales. As a result, effects of other energy scales get suppressed and the quasiparticle masses (both in the isotropic and anisotropic cases) are primarily governed by the temperature dependence of the thermal distributions. From the expressions for the quasiparticle masses in the high temperature regime, we have $m_{fT}\approx\mathcal{O}(gT)$ (eq. \eqref{Q.P.M.(Quark mass)}) and $m_{gT}\approx\mathcal{O}(gT)$ (eq. \eqref{Q.P.M.(Gluon mass)}) in the isotropic case, and $m_{fT\xi}\approx\mathcal{O}(gT)+\xi\mathcal{O}(gT)$ (eq. \eqref{Q.P.M.Q.(Anisotropic medium)}) and $m_{gT\xi}\approx\mathcal{O}(gT)+\xi\mathcal{O}(gT)$ (eq. \eqref{Q.P.M.G.(Anisotropic medium)}) in the anisotropic case with weak-momentum anisotropy. This can be understood from the fact that, in the high temperature regime, the exponential suppression factors $e^{\frac{\omega_f-\mu}{T}}$, $e^{\frac{\omega_f+\mu}{T}}$ and $e^{\frac{\omega_g}{T}}$ appearing in the distribution functions for quarks, antiquarks and gluons, respectively become weaker when $T$ rises. Consequently, the quasiparticle masses ($m_{fT}$, $m_{gT}$, $m_{fT\xi}$ and $m_{gT\xi}$) exhibit an approximately linear temperature dependence over the 
considered temperature range. It is important to note that, this linear temperature dependence of the quasiparticle masses is not intended to represent the full nonlinear behavior expected from QCD at 
all temperatures, but rather an effective parametrization that accurately captures the dominant thermal behavior in the temperature range relevant to our analysis. In this region, the variation of the coupling constant (and hence the quasiparticle masses) with the temperature is relatively moderate, so a linear approximation in $T$ remains well justified and physically meaningful. Some quasiparticle frameworks incorporating a running coupling linked to the lattice QCD equation of state, indeed predict 
a nonlinear temperature dependence of the quasiparticles, but such detailed modeling lies beyond the scope and purpose of the present work. Our focus is on elucidating medium-induced effects on the transport coefficients and observables within a transparent and analytically tractable framework, where the adopted linear behavior is both reasonable and sufficient for the intended comparisons. 

This alteration of quasiparticle masses creates a noticeable difference between the dispersion relations in the isotropic (eq. \eqref{I.D.R.(Quark)} and eq. \eqref{I.D.R.(Gluon)}) and anisotropic (eq. \eqref{A.D.R.(Quark)} and eq. \eqref{A.D.R.(Gluon)}) media. The modification of the thermal masses of partons in an anisotropic environment transpires the behavior of their distribution functions in such environment. The distribution functions involve the modified masses of the partons. Therefore, the distribution functions in the isotropic medium use the $T$ and $\mu$-dependent quasiparticle masses (eq. \eqref{Q.P.M.(Quark mass)} and eq. \eqref{Q.P.M.(Gluon mass)}), whereas the distribution functions in the expansion-induced anisotropic medium use the $T$, $\mu$ and $\xi$-dependent quasiparticle masses (eq. \eqref{Q.P.M.Q.(Anisotropic medium)} and eq. \eqref{Q.P.M.G.(Anisotropic medium)}). Since the quasiparticle masses of both quark and gluon are significantly modulated by the anisotropy of the medium, the particle distribution functions may illustrate appreciable alteration due to the emergence of anisotropy. In figures \ref{Fig.tdf}, \ref{Fig.cpdf} and \ref{Fig.pdf}, quark and gluon distribution functions have been plotted as functions of the temperature, chemical potential and momentum, respectively at different values of anisotropic parameter and they are also compared with their isotropic counterparts. Figure \ref{Fig.tdf} describes that, above and below a certain temperature, the effects of anisotropy on the parton distribution functions are slightly opposite. For example, a decreasing effect on the quark distribution function is noticed for $\xi=0.2$ below $T \approx 0.34$ GeV and for $\xi=0.4$ below $T \approx 0.36$ GeV, whereas an increasing effect is found above these temperatures for corresponding cases. Similarly, for gluon distribution function, these temperatures are $T \approx 0.37$ GeV for $\xi=0.2$ and $T \approx 0.40$ GeV for $\xi=0.4$. However, no such turning points along the variations of parton distribution functions with chemical potential (figure \ref{Fig.cpdf}) and with momentum (figure \ref{Fig.pdf}) are found, rather, increasing effect on the parton distribution functions due to the anisotropy is observed over the entire ranges of chemical potential and momentum. Thus, from the observation on the distribution functions, it is inferred that the parton number densities get enhanced with the increase of anisotropy. The effect of anisotropy on parton distribution functions is more evident at high temperatures (figure \ref{Fig.tdf}) and low momenta (figure \ref{Fig.pdf}), whereas the anisotropic impact remains nearly uniform throughout the considered range of chemical potential (figure \ref{Fig.cpdf}). These observations on parton distribution functions facilitate the understanding of the impact of expansion-induced anisotropy on various transport coefficients and observables for a baryon asymmetric hot QCD medium. 

\section{Results and discussions}
\begin{figure}[]
\begin{center}
\begin{tabular}{c c}
\includegraphics[width=7.4cm]{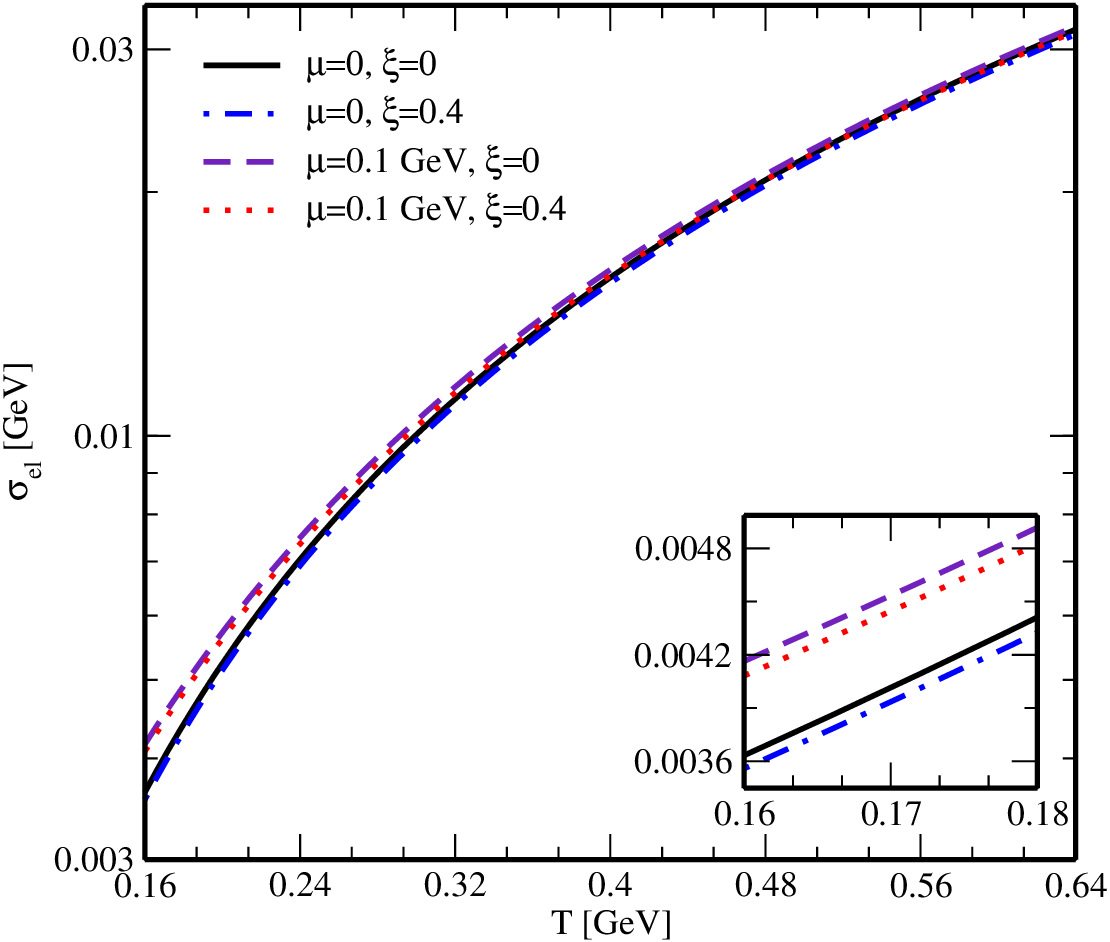}&
\hspace{0.73 cm}
\includegraphics[width=7.4cm]{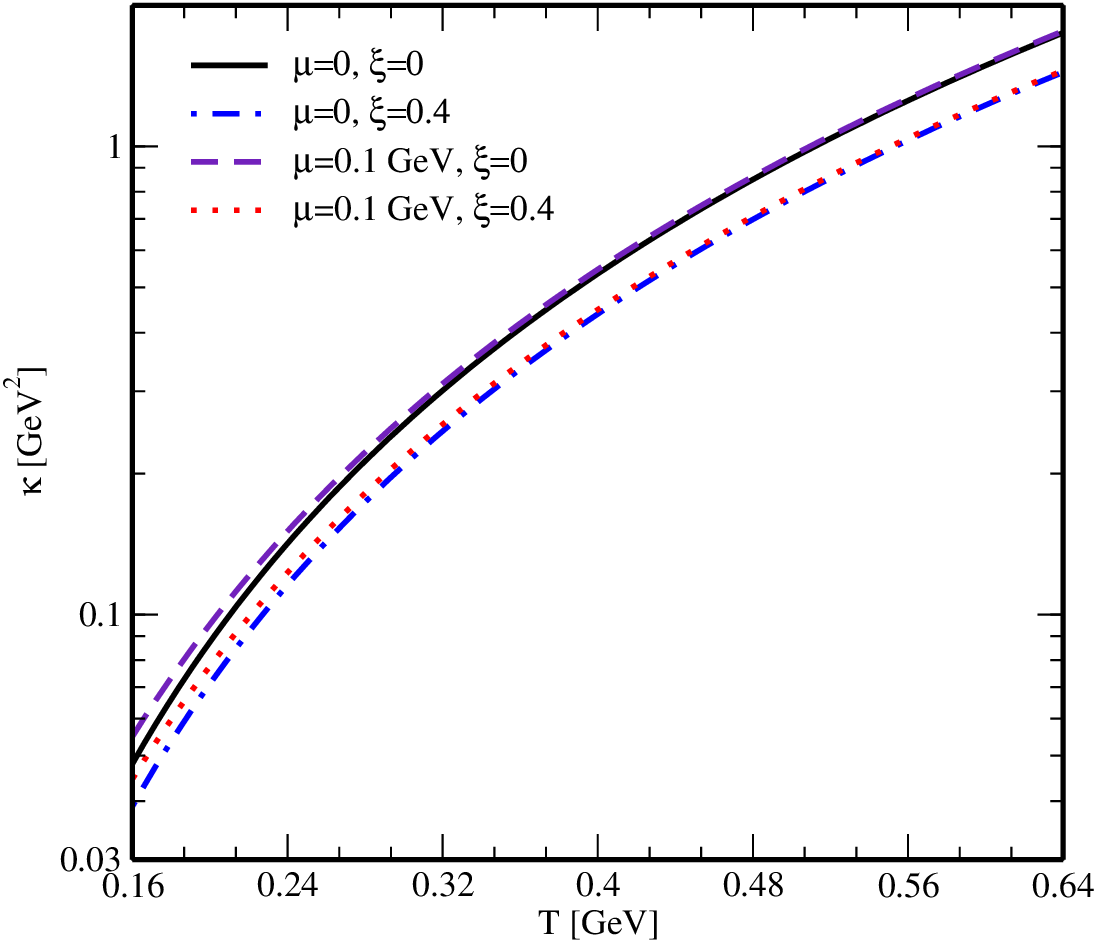} \\
a & b
\end{tabular}
\caption{Variations of (a) the electrical conductivity and (b) the thermal conductivity with temperature for different values of anisotropic parameter and chemical potential.}\label{Fig.1}
\end{center}
\end{figure}

\begin{figure}[]
\begin{center}
\begin{tabular}{c c}
\includegraphics[width=7.4cm]{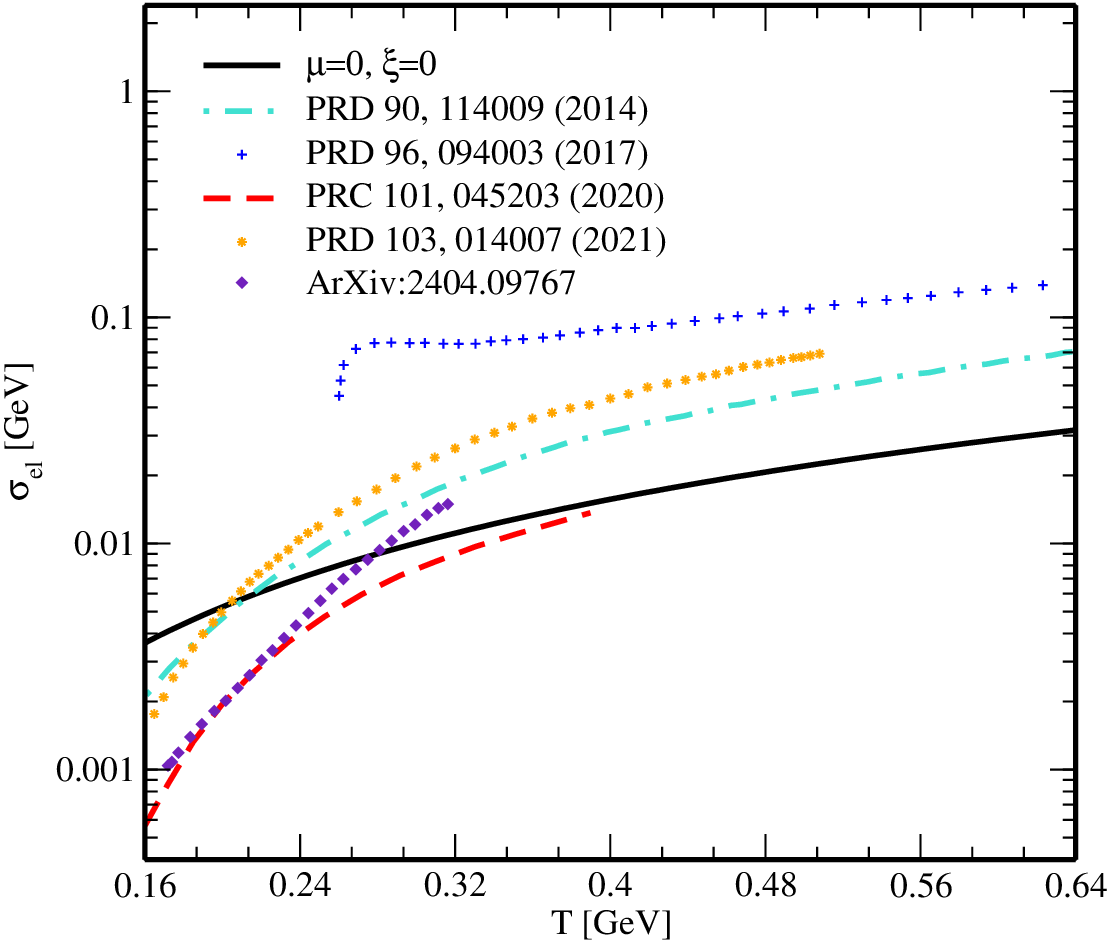}&
\hspace{0.73 cm}
\includegraphics[width=7.4cm]{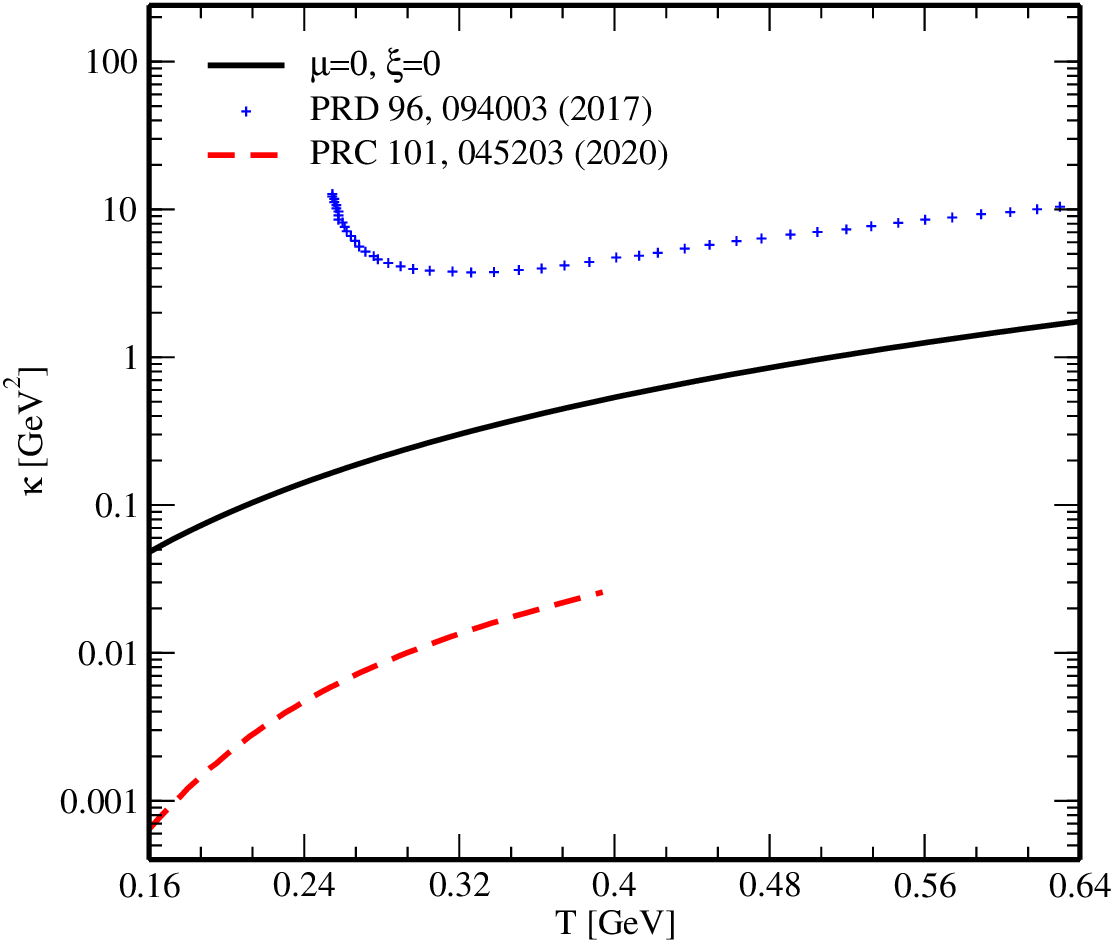} \\
a & b
\end{tabular}
\caption{Variations of (a) the electrical conductivity and (b) the thermal conductivity with the temperature for a baryonless isotropic medium and their comparison with the results of other works.}\label{Fig.Isotropy}
\end{center}
\end{figure}

In figures \ref{Fig.1}a and \ref{Fig.1}b, the electrical conductivity ($\ec$) and the thermal conductivity ($\kappa$) are respectively plotted with the temperature for different conditions of anisotropy and chemical potential. An increasing trend of $\ec$ and $\kappa$ with temperature is noticed for all conditions. It is further observed that the emergence of expansion-induced anisotropy decreases both $\ec$ and $\kappa$ for baryonless isotropic medium as well as for baryon asymmetric isotropic medium, thus indicating a reduction 
in the charge and heat conduction when the isotropic medium gets exposed to the expansion-induced 
anisotropy. Decrease of both $\ec$ and $\kappa$ can be perceived from their expressions (equations \eqref{A.E.C.(1)} and \eqref{A.T.C.(1)}), where their corresponding anisotropic parts ($\ec^\xi$ and $\kappa^\xi$) contribute in their decreasing trend, because these parts explicitly depend on the anisotropic parameter ($\xi$) and implicitly depend on $\xi$ through the dispersion relations and distribution functions. Since quasiparticle masses and distribution functions of partons increase with anisotropy, an enhancement 
in the negative magnitudes of $\ec^\xi$ and $\kappa^\xi$ is observed. Thus, it results in an overall reduction of the total electrical conductivity and the total thermal conductivity. However, the baryon asymmetric medium estimates larger values of $\ec$ and $\kappa$ as compared to their counterparts in the baryonless medium for isotropic scenario as well as for anisotropic scenario. The influence of baryon asymmetry on the conduction of charge and heat is more pronounced in the low temperature regime than that in the high temperature regime. This increase in the aforesaid conductivities at finite chemical potential is primarily attributed to the enhanced parton number densities in the baryon asymmetric medium. Thus the chemical potential plays a key role in enabling the flow of charge and heat in the medium, irrespective of whether the medium exhibits isotropy or anisotropy. 

Figure \ref{Fig.Isotropy} shows the comparison of our results on the electrical and the thermal conductivities at $\mu=0$ and $\xi=0$ with the corresponding results in different quasiparticle 
models \cite{Puglisi:PRD90'2014,Mitra:PRD96'2017,Soloveva:PRC101'2020,Mykhaylova:PRD103'2021,arXiv:2404.09767}. According to figure \ref{Fig.Isotropy}a, our $\sigma_{\rm el}$ agrees closely with the estimations of references \cite{Puglisi:PRD90'2014,Mykhaylova:PRD103'2021} at lower temperatures and with the estimations of references \cite{Soloveva:PRC101'2020,arXiv:2404.09767} at higher temperatures, whereas ref. \cite{Mitra:PRD96'2017} yields consistently higher values throughout the high temperature range. Figure \ref{Fig.Isotropy}b shows that, our $\kappa$ lies below the estimation of ref. \cite{Mitra:PRD96'2017} at high temperatures and above that of ref. \cite{Soloveva:PRC101'2020} at low temperatures. These quantitative differences arise from distinct modeling choices in each approach. For example, ref. \cite{Puglisi:PRD90'2014} uses the relaxation time approximation and the quasiparticle model in determining $\sigma_{\rm el}$ with the relaxation times derived from the transport cross sections, ref. \cite{Mitra:PRD96'2017} employs the effective fugacity quasiparticle model in exploring the transport coefficients, ref. \cite{Soloveva:PRC101'2020} computes the transport coefficients within the dynamical quasiparticle model using the temperature and baryon chemical potential-dependent parton interaction rates, ref. \cite{Mykhaylova:PRD103'2021} determines $\sigma_{\rm el}$ in a quasiparticle framework, where the interactions with the hot medium are embodied in effective masses of the constituents through a temperature-dependent running coupling extracted from the lattice QCD thermodynamics, and ref. \cite{arXiv:2404.09767} incorporates the nonperturbative resummation via Gribov gluon propagator in the quasiparticle approach and uses the relaxation times calculated from microscopic scattering amplitudes in determining the electrical conductivity. These methodological differences naturally lead to the variations observed in figure \ref{Fig.Isotropy}. 

\begin{figure}[]
\begin{center}
\begin{tabular}{c c}
\includegraphics[width=7.4cm]{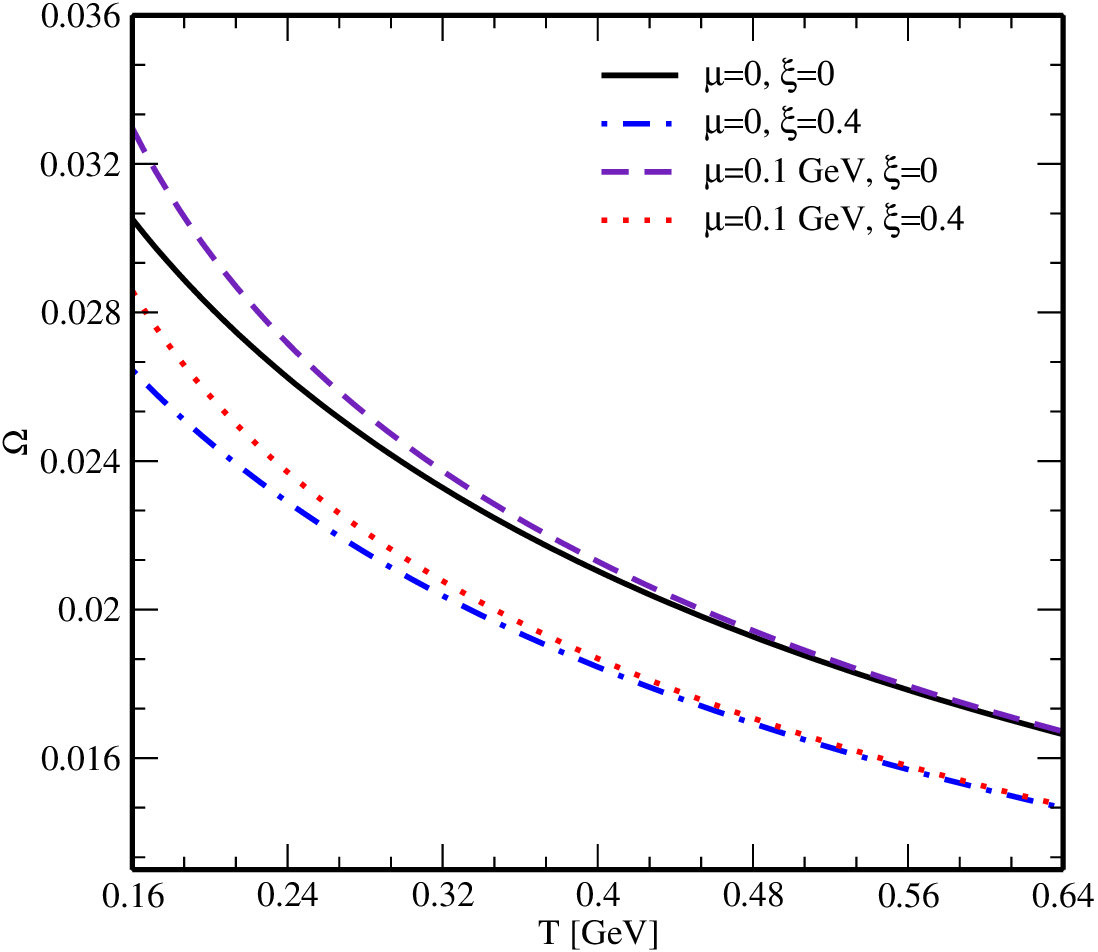}&
\hspace{0.73 cm}
\includegraphics[width=7.4cm]{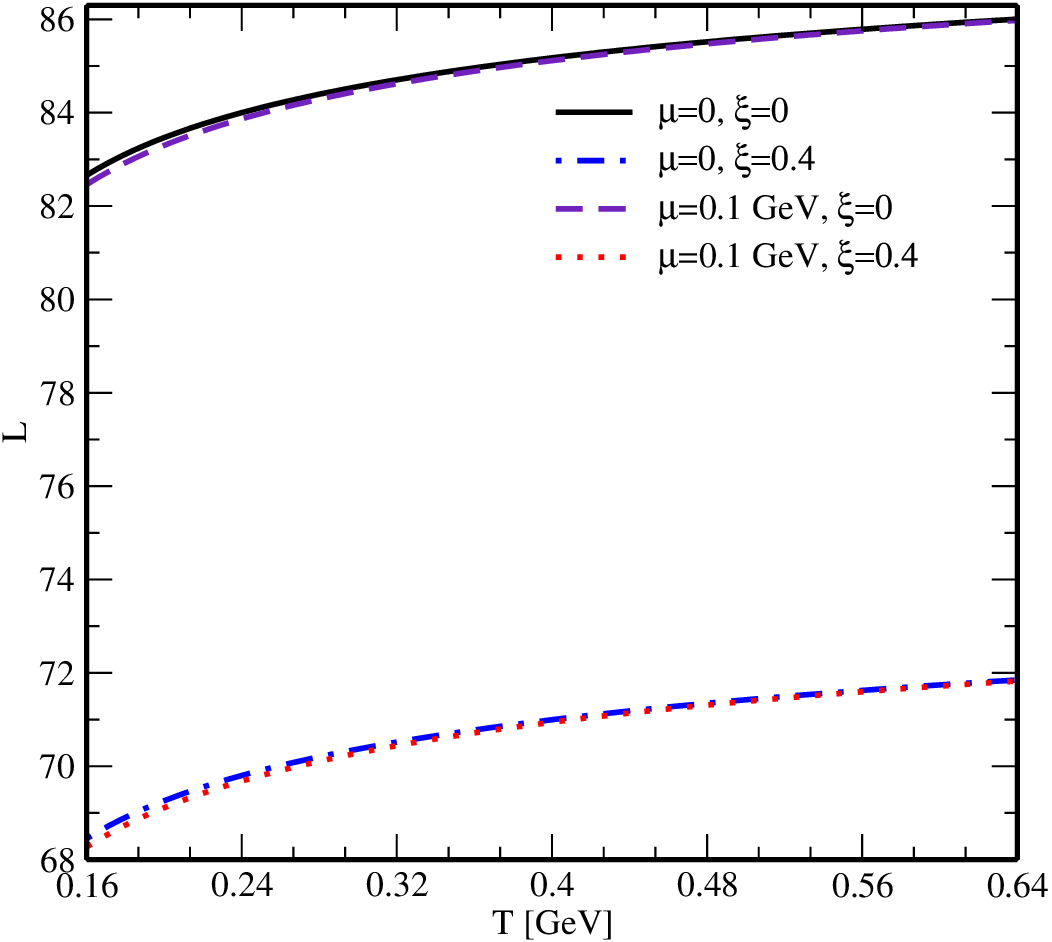} \\
a & b
\end{tabular}
\caption{Variations of (a) the Knudsen number and (b) the Lorenz number with temperature for different values of anisotropic parameter and chemical potential.}\label{Fig.2}
\end{center}
\end{figure}

Figure \ref{Fig.2}a displays the variation of the Knudsen number with the temperature for four different media, such as the baryonless isotropic medium, the baryonless anisotropic medium, the isotropic medium with finite baryon asymmetry and the anisotropic medium with finite baryon asymmetry. In all these 
media, the Knudsen number decreases with the increasing temperature, explaining that the mean free path becomes much smaller than the characteristic length of the medium as the medium becomes hotter. Thus, these media approach their corresponding equilibrium states faster as the temperature rises. When a baryonless medium exhibits an anisotropy due to the asymptotic expansion, the magnitude of the Knudsen number gets decreased and this deviation from its value in the isotropic medium is considerable. For baryon asymmetric medium also, a similar deviation is observed when the expansion-induced anisotropy is introduced. As a result, the emergence of anisotropy pushes the medium towards the equilibrium state. However, the magnitude of the Knudsen number is found to increase with the introduction of baryon asymmetry and this deviation from its counterpart in a baryonless medium is meagre for isotropic scenario as well as for anisotropic scenario. It indicates that the baryon asymmetry takes the medium a bit away from the equilibrium state. Thus, overall, the expansion-induced anisotropy and the baryon asymmetry leave conspicuous imprints on the equilibrium property of the medium. Since the decreasing behavior of the Knudsen number with the temperature holds for all curves, the primary distinctions in the finite $\mu$ and the finite $\xi$ cases lie in the magnitude rather than the qualitative temperature dependence. Among the various types of media, the isotropic medium with finite baryon asymmetry exhibits the highest value of the Knudsen number, while the baryonless medium with expansion-induced anisotropy yields the lowest value of the Knudsen number. Consequently, the likelihood of the system being in local equilibrium is minimal in the former and maximal in the latter. 

In figure \ref{Fig.2}b, the variation of the Lorenz number with the temperature for the four previously discussed media has been illustrated. It is observed that the Lorenz number gets decreased due to the 
emergence of expansion-induced anisotropy. Additionally, the medium with finite baryon asymmetry estimates lower value of the Lorenz number as compared to that in a baryonless medium and this effect of baryon asymmetry on the Lorenz number is only noticeable at low temperatures. Overall, the effect of expansion-induced anisotropy on the Lorenz number is more pronounced than that of baryon asymmetry. An increase in the temperature of the medium leads to a rise in the Lorenz number in all scenarios, thereby signaling a violation of the Wiedemann-Franz law. However, the Lorenz number in the abovementioned media remains greater than unity, inferring that the thermal conduction prevails over the charge conduction within the considered temperature range. The dominance of the thermal conductivity over the electrical conductivity is more evident in an isotropic medium than that in an anisotropic medium. 

\section{Conclusions}
In this work, we studied the effect of expansion-induced anisotropy on the electrical conductivity, the thermal conductivity, the Knudsen number and the Lorenz number of a baryon asymmetric hot QCD medium. In 
calculating the aforesaid conductivities, we solved the relativistic Boltzmann transport equation 
in the relaxation time approximation within the kinetic theory approach, where the interactions among partons were incorporated through their distribution functions in the quasiparticle approach at finite temperature, anisotropy and baryon asymmetry. Additionally, we determined the quasiparticle masses of partons for a baryon asymmetric hot QCD medium in the presence of expansion-induced anisotropy and noticed an increasing trend 
of the quasiparticle masses of partons with the increase of anisotropy. This led to an enhancement 
in the parton number densities in the anisotropic medium when compared to the isotropic case. We also 
compared the aforesaid transport coefficients and observables with their counterparts in the isotropic medium as well as in the baryonless medium. Our observation found that the introduction of anisotropy tends to 
reduce the conduction of charge and heat in the hot QCD medium. On the other hand, the emergence of baryon asymmetry enhances the aforesaid conductivities, thus facilitating the charge and heat transport phenomena 
in the medium. The anisotropy and the baryon asymmetry also left significant imprints on the observables associated with the aforesaid conductivities, such as the Knudsen number and the Lorenz number of the medium. We found that the expansion-induced anisotropy decreases the Knudsen number, contrary to its increase 
due to the baryon asymmetry, thus discerning the fact that the medium is closer to the local equilibrium state in the presence of abovementioned anisotropy. Further, a significant reduction in the Lorenz number was observed due to expansion-induced anisotropy, while baryon asymmetry caused only a marginal decrease. Consequently, the impact of expansion-induced anisotropy on the Lorenz number is more pronounced than that of baryon asymmetry. This observation on the Lorenz number signifies that the dominance of the heat transport 
over the charge transport gets reduced in the expansion-induced anisotropic medium as well as in the baryon asymmetric medium. In the low temperature regime, the Lorenz number exhibited an increasing trend 
with the temperature, indicating a violation of the Wiedemann-Franz law. 

The observations in this work on the electrical and the thermal conductivities in the presence of expansion-induced anisotropy for a baryon asymmetric hot QCD medium hold significant phenomenological importance. The dilepton production rate was noted to be directly proportional to the electrical conductivity of the quark-gluon plasma \cite{Moore:arXiv0607172}. Thus, the reduction in electrical conductivity caused by expansion-induced anisotropy may suggest a lower dilepton production rate in experiments at ultrarelativistic heavy 
ion collisions. On the other hand, the increase of electrical conductivity due to baryon asymmetry may indicate a higher dilepton production rate. Additionally, it was suggested that the anisotropy of electrical conductivity facilitates the generation of elliptic flow of photons \cite{Yin:PRC90'2014}. Our result also exhibits anisotropy due to the asymptotic expansion of matter, so it may have a significant impact on the elliptic flow of photons. The thermal conductivity, being linked to the local equilibrium of the 
medium through the mean free path, can be used to gauge how close the medium produced in ultrarelativistic heavy ion collisions is to the equilibrium state in the presence of expansion-induced anisotropy. The exploration of the electrical and the thermal conductivities in a baryon asymmetric hot QCD medium with expansion-induced anisotropy is valuable for understanding conductive properties in other areas where anisotropy and baryon asymmetry may be present, such as the cores of dense magnetars and the beginning 
of the universe, in addition to ultrarelativistic heavy ion collisions. 

\section{Acknowledgments}
The author acknowledges financial support from ANID Fondecyt Postdoctoral Grant 3240349 for this work. The author would like to thank Nicol\'{a}s A. Neill for helpful discussions.

\end{document}